\documentclass{aastex}
\usepackage{graphicx,epsfig,emulateapj5}

\newcommand{\etal}{{et al.~}}

\newcommand{\bq}{\begin{equation}}
\newcommand{\eq}{\end{equation}}

\def\gtsim{\lower.5ex\hbox{$\buildrel > \over\sim$}}
\def\ltsim{\lower.5ex\hbox{$\buildrel < \over\sim$}}
\def\arcsec{^{\prime\prime}}
\def\arcmin{^\prime}

\def\sun{\mbox{$_\odot$}}
\def\apjl{ApJL}
\def\apj{ApJ}
\def\apjs{ApJS}
\def\mnras{MNRAS}
\def\araa{ARAA}
\def\aj{AJ}
\def\aap{A\&A}
\def\aaps{A\&A Suppl.}
\def\nat{Nature}
\def\pasp{PASP}
\slugcomment{Submitted to  The Astrophysical Journal} 
\shorttitle{Barred Galaxies in the Abell 901/2 supercluster}
\shortauthors{Marinova, Jogee, Heiderman, \& the STAGES collaboration}

\begin{document}
\title
{Barred Galaxies in the Abell 901/2 Supercluster with STAGES}
\author {Irina Marinova\altaffilmark{1}, Shardha
  Jogee\altaffilmark{1}, Amanda Heiderman\altaffilmark{1}, Fabio
 D. Barazza\altaffilmark{3}, Meghan
 E. Gray\altaffilmark{2},  Marco Barden\altaffilmark{9}, Christian Wolf\altaffilmark{14},
Chien Y. Peng\altaffilmark{5,6}, David Bacon\altaffilmark{7}, Michael
 Balogh\altaffilmark{8}, Eric F. Bell\altaffilmark{4}, Asmus
 B\"ohm\altaffilmark{9,10}, John A.R. Caldwell\altaffilmark{16}, Boris H\"au\ss ler\altaffilmark{2},
 Catherine Heymans\altaffilmark{12}, Knud Jahnke\altaffilmark{4},
 Eelco van Kampen\altaffilmark{9}, Kyle Lane\altaffilmark{2}, Daniel H. McIntosh\altaffilmark{13,17}, Klaus Meisenheimer\altaffilmark{4}, Sebastian
 F. S\'anchez\altaffilmark{11}, Rachel Somerville\altaffilmark{4},
 Andy Taylor\altaffilmark{12}, Lutz Wisotzki\altaffilmark{10}, $\&$ Xianzhong Zheng\altaffilmark{15}
 }
\authoremail{marinova@astro.as.utexas.edu, sj@astro.as.utexas.edu}
\altaffiltext{1}{Department of Astronomy, University of Texas at
Austin, Austin, TX } 
\altaffiltext{2}{School of Physics and Astronomy, The University of
Nottingham, Nottingham, UK} 
\altaffiltext{3}{Laboratoire
d'Astrophysique, \'Ecole Polytechnique F\'ed\'eralede Lausanne,
Observatoire, Sauverny, Switzerland} 
\altaffiltext{4}{Max-Planck-Institut f\"{u}r Astronomie, Heidelberg, Germany} 
\altaffiltext{5}{NRC Herzberg
Institute of Astrophysics, Victoria,Canada} 
\altaffiltext{6}{Space
Telescope Science Institute, Baltimore, MD, USA}
\altaffiltext{7}{Institute of Cosmology and Gravitation, University of
Portsmouth, Portsmouth, UK} 
\altaffiltext{8}{Department of Physics
and Astronomy, University Of Waterloo, Ontario, Canada}
\altaffiltext{9}{Institute for Astro- and Particle Physics,
University of Innsbruck, Innsbruck, Austria}
\altaffiltext{10}{Astrophysikalisches Insitut Potsdam, Potsdam,
Germany} 
\altaffiltext{11}{Centro Hispano Aleman de Calar Alto,
Almeria, Spain} 
\altaffiltext{12}{The Scottish Universities Physics
Alliance, Institute for Astronomy, University of Edinburgh, Edinburgh,
UK} 
\altaffiltext{13}{Department of Astronomy, University of
Massachusetts, Amherst, MA, USA} 
\altaffiltext{14}{Department of
Astrophysics, University of Oxford, Oxford, UK}
\altaffiltext{15}{Purple Mountain Observatory, National Astronomical
Observatories, Chinese Academy of Sciences, Nanjing, China}
\altaffiltext{16}{University of Texas, McDonald Observatory, Fort
Davis, TX, USA}
\altaffiltext{17}{Department of Physics, University of Missouri-Kansas City, Kansas City, MO
64110, USA}

\begin{abstract}
We present a study of bar and host disk evolution in a dense cluster environment, 
based on a sample of $\sim$~800 bright ($M_{\rm V} \le -18$) galaxies in the Abell 901/2  
supercluster at $z\sim$~0.165. 
We use $HST$ ACS F606W imaging from the STAGES survey, and data from $Spitzer$, $XMM-Newton$,
and COMBO-17.
We identify and characterize bars through ellipse-fitting, and other morphological
features through visual classification. We find the following results: 
(1)~To define the optical fraction of barred disk galaxies, 
we explore three commonly used methods for selecting disk galaxies.  
We find 625, 485, and 353 disk galaxies, respectively, via visual 
classification, a single component S{\'e}rsic cut ($n\le2.5$), and  a blue-cloud cut. 
In cluster environments, the  latter two methods suffer from  serious limitations, 
and  miss 31\% and 51\%, respectively, of visually-identified disks, particularly the many 
red, bulge-dominated disk galaxies in clusters. 
(2)~For moderately inclined disks, the three methods of disk selection, 
however, yield a similar global optical bar fraction 
($f_{\rm bar-opt}$) of 34\%$^{+10\%}_{-3\%}$ (115/340), 31\%$^{+10\%}_{-3\%}$ (58/189),  
and 30\%$^{+10\%}_{-3\%}$ (72/241), respectively.  
(3)~We explore  $f_{\rm bar-opt}$ as a function of host galaxy properties and find that 
it rises in brighter galaxies and those which appear to have no significant bulge component.
Within a given  absolute magnitude  bin,  $f_{\rm bar-opt}$ is higher in visually-selected 
disk galaxies that have no bulge as opposed to those with bulges. 
Conversely, for a given visual morphological class,  $f_{\rm bar-opt}$ 
rises at higher luminosities. Both results are similar to trends found in the field.
(4) For bright early-types, as well as faint late-type systems with no evident bulge, 
the optical bar fraction in the Abell 901/2 clusters is comparable within a factor of 1.1 
to 1.4  to that of  field galaxies at lower redshifts ($z<0.04$).
(5)~Between the core and the virial radius of the cluster ($R\sim$~0.25 to 1.2 Mpc) at 
intermediate environmental densities (log($\Sigma_{10}$)$\sim$~1.7 to 2.3), 
the optical bar fraction does not appear to depend strongly on the local environment density 
tracers  ($\kappa$, $\Sigma_{10}$, and ICM density), and varies at most by a factor of 
$\sim$~1.3. Inside the cluster core, we are limited by number statistics, projection
effects, and different trends from different indicators, but overall 
$f_{\rm bar-opt}$ does not show evidence for a variation larger than a factor of 1.5. 
We discuss the implications of our results 
for the evolution of bars and disks in dense environments.
\end {abstract}


\vskip 0.5 in 
\section{Introduction}\label{intro}

Stellar bars are one of the most important internal drivers of disk
galaxy evolution.  For field galaxies in the local universe, 
bars are known to be the most efficient way to redistribute material in the 
galaxy disk (Combes \& Sanders 1981; Weinberg 1985; Debattista \& Sellwood 1998, 2000;
Athanassoula 2002). Bars channel gas into the central regions of galaxies, where
powerful starbursts can ignite (Schwarz 1981; Shlosman, Frank, \& Begelman 1989;
 Kormendy \& Kennicutt 2004; Jogee 1999; Jogee, Scoville, \& Kenney
2005; Sheth et al. 2005), building central disky structures known as
`pseudobulges' (Kormendy 1982; Kormendy 1993; Jogee 1999; Jogee, Scoville, \&
Kenney 2005; Fisher 2006; Weinzirl et al. 2009). 
Peanut/boxy bulges in inclined
galaxies are thought to be associated with bending instabilities and vertical  
resonances in bars (e.g., Combes \& Sanders 1981; Combes \etal 1990; Pfenniger  \& Norman 1990;  
Athanassoula 2005; Martinez-Valpuesta \etal 2006).

As early as 1963, de Vaucouleurs used
visual classification on photographic plates to find that
approximately 30\% of nearby galaxies appear strongly barred in the
optical band, with the fraction increasing to approximately 
60\% if very  weak bars are considered. 
Quantitative studies for the optical bar fraction at $z\sim$~0 
yield a mean value of 45\% to 52\% with a typical uncertainty of $+8\%$ 
from ellipse-fits  (Marinova \& Jogee 2007, hereafter MJ07; 
Barazza, Jogee, \& Marinova 2008; hereafter BJM08;  Aguerri et
al. 2009, hereafter A09) and $\sim$~47\% from  bulge-disk-bar decomposition  (Reese et al. 2007). The lower value 
from these quantitative methods  compared to the 60\%  value from de Vaucouleurs (1963)
stems from the fact that many weak bars (with RC3 class `AB')  are obscured 
by dust and star formation (SF), caused by the presence of  curved shocks/dust lanes  (e.g., Athanassoula
1992) on the leading edges of the bar. Many such bars may fail to 
meet rigorous quantitative criteria for characterizing bars via ellipse-fit or 
bulge-disk-bar decomposition, but their presence can sometimes be guessed via 
visual inspection (see MJ07 for detailed discussion).  
In the  near infra-red (NIR), where obscuration by dust and SF is minimized,  
different quantitative methods, such as  ellipse-fit  (Menendez-Delmestre et al. 2007; MJ07), 
bulge-disk-bar decomposition  (Weinzirl et al. 2009)  and Fourier decomposition  
(Laurikainen et al.  2004) all yield a NIR bar fraction of $\sim$~60\% for bright nearby
samples.

The above values of the bar fraction at $z\sim$~0  refer to the globally-averaged 
value over a wide range of Hubble types and luminosities. Several studies have 
performed more detailed explorations to  look at how bars relate to the properties 
of the host spiral galaxies. 
Recent studies based on SDSS  (BJM08;  A09) using ellipse-fits 
report that  the optical bar fraction  rises in spiral galaxies which
appear to be disk-dominated, 
quasi-bulgeless, or have a morphology suggestive of a low  
bulge-to-disk ratio. A similar trend was observed  by Odewahn (1996) using 
visual classes: he found that the optical fraction of strong bars in disk galaxies 
rises from Sc galaxies towards later types. 
Similar results are found in the near-infrared by Weinzirl et
al. (2008) using 2D bulge-disk-bar decomposition on nearby bright 
spiral galaxies.

Recently, studies performed at intermediate redshifts with the Advanced Camera for
Surveys (ACS) on the Hubble Space Telescope (HST) have allowed bars to be probed 
at earlier epochs. Several studies have shown that the optical fraction of strong
($e>0.4$) or prominent bars is $\sim$~30\% on average over  $z\sim$~0.2--1  
(Elmegreen et al. 2004; Jogee et al. 2004; Zheng et al. 2005).  In particular,  
Jogee et al. (2004) find that the 
optical fraction of strong  bars  does not show an order of magnitude decline, 
but only varies  from 36\%$\pm$6\% over $z\sim$~0.2--0.7 
to 24\%$\pm$4\%  over $z\sim$~0.7 to 1.0.  A  much larger study  finds a 
variation in  the optical fraction of strong bars from  27\%$\pm$1\%  to  
12\%$\pm$1\% for $z\sim$~0.2--0.84 (Sheth et al. 2008).  Interpretations differ  on whether the observed  decline 
is simply due to systematic effects, such as loss of resolution and rising obscuration  with
redshift (Jogee et  al. 2004; MJ07;  BJM08), or whether it reflects an intrinsic 
decline  (Sheth et al. 2008) in the true bar fraction.

While bars have been studied extensively in the field, little is known about the 
fraction of bars and their properties in dense environments. 
The presence of bars is particularly useful to identify galaxies with disks 
in clusters ($\S$~\ref{cldisks}), where other disk signatures, such as spiral arms, 
may be absent due to ram pressure stripping.
Furthermore, we can use galaxy clusters as a lab to test our theories of bar formation 
and evolution.  The fraction of barred galaxies in a cluster depends on the 
epoch of bar formation, the robustness of bars, the interplay between cluster
environmental processes (harassment, tidal interactions, ram pressure stripping), 
and the evolutionary history of clusters.

The detailed process of bar formation is not yet known, but simulations
suggest that a cold disk, with low velocity dispersion, $\sigma$,  favors the formation of
spontaneous disk instabilities  (e.g., Toomre 1964; Jog \& Solomon 1984). External triggers, such as tidal 
interactions, can also induce bars in a dynamically cold disk 
(e.g., Noguchi 1987; Elmegreen et al. 1990; Elmegreen et al. 1991; 
Hernquist \& Mihos 1995). Thus, cluster processes can have competing effects on bar formation. 
While frequent tidal interactions can induce stellar bars, they may also heat
the disks and thereby make them less susceptible to bar formation. Dubinski et al. (2008) 
explored these effects by modeling the interaction of a hundred DM satellites on 
M31. They found that while the satellites did not have a large heating effect on 
the disk, encounters close to the galaxy center could produce strong non-axisymmetric 
instabilities such as stellar bars. However, in dense clusters, disk galaxies that
are deprived of their cold gas through ram-pressure stripping may be too dynamically 
hot to form bars. Recently van den Bosch et al. (2008) have shown that low-mass satellite 
cluster galaxies may be more affected by gas strangulation, which may in turn make
them less favorable to bar formation. 
It is also important to note that if bars cannot be easily dissolved once
formed (see $\S$~\ref{disc}), then in scenarios where clusters grow by 
accretion of field galaxies, existing bars in accreted galaxies may not be much impacted by 
subsequent cluster processes.

There have been only a handful of observational studies that have explored the impact 
of environment on  barred disks. Recently, A09 studied the effects of 
environment in field and intermediate density regions on barred galaxies using $\sim$~3000 galaxies at $0.01 \le z \le 0.04$ 
from SDSS-DR5, and found that the bar fraction and properties were not correlated to 
galaxy environment. Bars were identified using ellipse fits. 
However, they excluded interacting galaxies from their study.  Barazza et al. 
(2009) study the impact of environment on bars in disk galaxies using $\sim$~2000 galaxies 
at intermediate redshift  ($z\sim$~0.4--1) from the  ESO Distant Clusters Survey 
(EDisCS; White et al. 2005). van den Bergh (2002) found no difference between the bar
fraction in the field and in clusters using a uniform sample of 930 galaxies from the
Shapley-Ames catalog, in a study where bar classifications were performed
through visual inspection of optical images. He therefore concluded that
the bar fraction depends solely on host-galaxy properties. It should
be noted that for this study, the environment assignments were largely
 qualitative, made by inspecting the region around the
galaxy on the image, and looking at luminosities and radial velocities
of surrounding galaxies. Varela et al. (2004)
found that the bar fraction is almost twice as high in galaxies that
are interacting, compared to isolated galaxies. Their study 
relied on redshifts from the CfA survey and morphological classifications 
from LEDA and NED. 
The results of Varela et al. (2004) confirm previous studies (e.g., Elmegreen et al. 1990), who find a higher number of 
barred galaxies in binary systems.

We are now in a position to make further progress in this largely unexplored aspect of galaxy
evolution with the STAGES panchromatic dataset ($\S$~\ref{datsamp}), which includes: a 0.5
$\times$ 0.5 square degree HST ACS mosaic in F606W of the A901/2 supercluster,
spectrophotometric redshifts from COMBO-17, coverage with $XMM-Newton$,
$GALEX$, and $Spitzer$, as well as dark matter maps. In $\S$~\ref{method}, we
outline the techniques for characterizing bars and disks. It should be noted that traditionally the bar fraction
$f_{\rm bar-opt}$ is defined as the fraction of \textit{disk galaxies}
that are barred. Hence calculation of $f_{\rm bar-opt}$  
requires disk galaxies to be reliably identified. We use the term `disk galaxies' to describe all galaxies with an
outer disk component (e.g., S0-Sm), which may or may not be accompanied by a
central bulge. In this paper, we draw
attention to the fact that many automated methods commonly 
used to identify disks in the field may fail in clusters.
Motivated by this, we explore different ways of identifying disks 
(e.g., color cut, S{\'e}rsic cut, visual classification) in  $\S$~\ref{cldisks}, 
and explore the effect on $f_{\rm bar-opt}$. We determine 
the frequency of bars as a function of host disk properties
($\S$~\ref{fbarBT}, \ref{fbarL}, \ref{fbarcolor}), and as 
a function of cluster radius, galaxy number density, ICM density, and
DM density ($\S$~\ref{fbardens}). The comparison of our results to those
from field studies is given in $\S$~\ref{fieldcomp}. In
$\S$~\ref{disc}, we discuss the implications
of our results for the evolution of bars and disks in 
dense environments. In $\S$~\ref{sum}, we give the summary
and conclusions.

\section{Data and Sample Selection}\label{datsamp}

The Abell 901/902 supercluster consists of three galaxy
clusters and a group at $z\sim$~0.165, with an average separation of 1~Mpc. 
The properties of this system are described in detail in Gray et
al. (2002).  The STAGES survey (Gray et al. 2008) covers a 0.5 $\times$ 0.5
square degree field centered on the supercluster, consisting of an 80-tile mosaic with the HST ACS
F606W. This ACS filter corresponds closely to the optical $B$ band. 
The ACS point spread function (PSF) of 0.1$\arcsec$ corresponds to $\sim$~282 pc at
$z\sim$~0.165\footnote{We assume in this
paper a flat cosmology with $\Omega_M = 1 - \Omega_{\Lambda} = 0.3$
and $H_{\rm 0}$ =70~km~s$^{-1}$~Mpc$^{-1}$.}. Spectro-photometric redshifts are available for all
galaxies from COMBO-17 (Wolf et al. 2004; 2005) where
the photo-z accuracy of the sample used in this paper is $\delta z/(1+z)\sim$~0.01.
 The multi-wavelength dataset includes X-ray maps of the
ICM density from $XMM-Newton$, UV from $GALEX$, $Spitzer$ 24$\mu$
coverage, and
dark matter maps from weak lensing (Heymans et
al. 2008). Total star formation rates (SFRs) derived from UV and 
$Spitzer$ 24$\mu$ luminosities (Bell et al. 2005), as well as stellar masses (Borch et
al. 2006) are also available for this field. 

Cluster galaxies are selected using photometric redshifts (see Gray et al. 2008 for a
detailed description). 
This provides a sample of 1990 cluster galaxies. For this paper, we
focus on galaxies brighter than $M_{\rm V} \le -18$. We choose this
cutoff, because it tends to separate well the regimes where normal and
dwarf galaxies dominate on the luminosity functions of clusters
(Binggeli, Sandage, \& Tammann 1988). We do not consider dwarf galaxies in this study for two
reasons. Firstly, our resolution of $\sim$~282 pc may be insufficient in
many cases to reliably identify morphological structures such as bars
in smaller dwarf galaxies. Secondly, the contamination of
the sample by field galaxies at magnitudes fainter than $M_{\rm V} = -18$ becomes
significant. This leaves us with a sample of 785 bright ($M_{\rm V} \le
-18$), cluster galaxies. The field contamination for this sample is estimated
to be $\sim$~10\%, from the space density of field galaxies using the
same absolute magnitude cut (Wolf, Gray, Meisenheimer 2005).

\section{Methodology}\label{method}

\subsection{Methods for Selection of Disk Galaxies}\label{diskselmeth}

In all studies conducted to date (e.g., deVaucouleurs 1963; Sellwood \&
Wilkinson 1993; Eskridge et al. 2000; Knapen et al. 2000; Mulchaey \&
Regan 1997; Jogee et al. 2004; Laurikainen et al. 2004; Elmegreen et al. 2004; Zheng et
al. 2005; Buta et al. 2005; MJ07;
Men{\'e}ndez-Delmestre et al. 2007; BJM08; Sheth et al. 2008), 
the bar fraction, $f_{\rm bar}$ has been defined as the 
number of barred \textit{disk} galaxies divided by 
the total number of \textit{disk} galaxies:
\begin{equation}
f_{\rm bar} = \frac{N_{\rm barred}}{N_{\rm disk}} = \frac{N_{\rm
      barred}}{N_{\rm barred} + N_{\rm unbarred}}. 
\end{equation}
Note that in the above studies, as well as in this paper, we use the term `disk galaxies' to describe all galaxies with a
significant outer disk component (e.g., systems typically labeled as S0-Sm), which may or may not be accompanied by a
central bulge. The bar fraction is only quoted with disk galaxies in mind, because
bars are believed to be related to an $m=2$ instability in the disk component of
galaxies. Furthermore, if the bar fraction were calculated over \textit{all} galaxies, 
changes in the morphological distribution between disk and spheroidal
(e.g., E) galaxies would 
influence the bar fraction and make it hard to compare across different samples. 
In the local universe, for 
nearby galaxies, catalogs like the RC3 (deVaucouleurs et al. 1991) contain visual 
classifications of galaxy morphology, making it possible to select
a sample of disk galaxies for bar studies. In large surveys such
as the SDSS and GEMS, two quantitative methods have been used to 
pick out disk galaxies: (1)~using a blue-cloud color cut in color-magnitude space
(Jogee et al. 2004; BJM08)
and (2)~using a S{\'e}rsic index, $n$, from a single component fit to
isolate a sample of disk-dominated galaxies (Jogee et al. 2004; Bell
et al. 2004; Barden et al. 2005; Ravindranath et
al. 2004). 
In the color cut method, only blue cloud galaxies are selected
on a $U-V$ color-magnitude diagram. The S{\'e}rsic cut method involves 
selecting only galaxies with S{\'e}rsic index $n <
2.5$. This is motivated by the fact that a pure
disk has a S{\'e}rsic index of 1, while a deVaucouleurs profile typically used
to describe a spheroid has a S{\'e}rsic index of 4. Note that in Bell
et al. (2004), Barden et al. (2005) and Ravindranath et al. (2004),
the goal was to broadly separate early-type
(E/S0/Sa) galaxies, from late-type disk-dominated galaxies
(Sb-Sm). However, because bars can occur in all types of disk galaxies
from S0-Sm, we would like to explore how well such  S{\'e}rsic and color cuts work in our cluster
sample at separating spheroidal galaxies (Es) from disk galaxies as
defined above (e.g., S0-Sm).

Using a blue-cloud or S{\'e}rsic cut to pick out
disk-dominated galaxies works fairly well at isolating a disk-galaxy
sample in the field. However, these methods
can grossly fail in a cluster environment, where the galaxy populations are different than those in the
field. Gas stripping of spirals could quench their star formation and
make them look redder. These galaxies might then be missed by a color
cut. On the other hand, the prevalence of bulge-dominated S0-type
disk galaxies in clusters (Dressler 1980) could be missed by a S{\'e}rsic
cut.  For this reason, we
use a third method to pick out disk galaxies: visual classification.

We visually classify the whole sample and put galaxies into different
groups according to the galaxy morphology ($\S$~\ref{vcmeth}). 
A galaxy is identified as a disk galaxy if it exhibits the dynamical
signatures of disk instabilities such as a stellar bar and spiral arms. 
In the absence of such structure, disks are picked by 
an identifiable break between the bulge and
disk component either in the image itself and/or looking for a break 
between a steep inner profile and a slowly declining outer profile in 
an estimation of
the brightness profile with the Smithsonian Astrophysical Observatory
visualization tool $DS9$. Three classifiers (I.M., A.H., S.J.) completed
a training set of several hundred galaxies, and two classifiers (A.H. \&
I.M.) classified the full cluster sample, with the third classifier
performing random checks. Subsequently, uncertain cases were
reviewed by all three classifiers. In our bright cluster galaxy sample of 785 galaxies, 
750 of them could be classified into visual classes as described above. 
The remaining galaxies were either too messy to classify, too
compact to classify, or unclassifiable for other reasons, such as noise
or edge effects. We could not reach
agreement on 4\% of cases regarding whether a galaxy was a pure bulge
or contained a disk component. 

From the three different methods of disk selection (visual, S{\'e}rsic cut, blue-cloud cut) we obtain
625, 485, and 353 disk galaxies, respectively. Detailed results from the different methods of
disk selection are presented in $\S$~\ref{cldisks}.

\subsection{Characterization of Bars}\label{barchar}

We use the standard IRAF task `ellipse' to fit ellipses to the galaxy
isophotes out to $a_{\rm max}$, where $a_{\rm max}$ is
the radius at which the surface brightness reaches sky level. 
This method of ellipse fitting has been widely used to identify
and characterize bars (e.g., Wozniak et al. 1995; Friedli et al. 1996;
Regan et al. 1997; Jogee et al. 1999, 2002a,
2002b, 2004; Knapen et al. 2000; Laine et al. 2002; Sheth et al. 2003;
Elmegreen et al. 2004; MJ07; Men{\'e}ndez-Delmestre et al. 2007). We employ
an iterative adaptive wrapper, developed by Jogee et al. (2004), which runs
the task `ellipse' up to a maximum number of $N$ iterations. Each
iteration uses the previous fit to produce an improved guess for the
isophote parameters. $N$ is typically set to 300, but for most objects
we obtain a good fit in only a few iterations. A good fit is one where
an ellipse is able to be fitted at every radial increment out to $a_{\rm max}$. 
As described in detail in Jedrzejewski (1987), the goodness of the ellipse fits is characterized 
by the harmonic amplitudes A3, B3, A4, and B4. The amplitudes of these
components signify how well the shape of the actual isophote is approximated by
the fitted ellipses (e.g., Jedrzejewski 1987).  For this sample, we find typical 
amplitudes of 0--15\% in the bar region. The advantages and limitations of the 
ellipse-fitting method are further discussed in 
detail in MJ07, where the statistical effects of deprojection are also addressed. 
We were able to successfully fit 97\% of the visually identified 
disk sample of 625 galaxies. The galaxies where `ellipse' fails generally do not have a
regularly decreasing surface brightness profile, which is necessary
to define the center for the fitting routine.  

We overlay the fitted ellipses onto the galaxy images and plot the
radial profiles of surface brightness (SB), ellipticity ($e$), and
position angle (PA). We use both the overlays and radial profiles to
classify the disk galaxies as `inclined',  `barred'  or `unbarred' 
using an interactive classification tool  and quantitative  criteria (Jogee et al. 
2004). The three classes are described below.  
We also extract quantitative parameters from the radial profiles, such as the 
size, ellipticity, and PA of both the disk and bar.

Disk galaxies classified as `inclined' have an outermost isophote with $e >0.5$, 
corresponding to $i > 60^{\circ}$. Because it is difficult to
identify morphological structures in such highly inclined disk galaxies, we
do not attempt to classify them as `barred' or `unbarred'. 
After discarding highly inclined disk galaxies (226 or 36\%) and those with visually-identified 
poor fits (32 or 5\%), we are left with 350 moderately inclined ($i < 60^{\circ}$), 
bright ($M_{\rm  V} \le -18$), cluster disk galaxies.  
The luminosity and color distributions of the total sample of 785 bright,
cluster galaxies, the visually-identified disk galaxy sample (N$=625$),  
and the moderately-inclined, ellipse-fitted sample of 350 disk galaxies
are over-plotted in Figure~\ref{MVcolordist}a and b. The figure shows that no 
significant bias is introduced on $M_{\rm  V}$ and color by restricting the sample to moderately inclined disk galaxies.

For galaxies with moderate inclinations ($i<60^{\circ}$), we classify a
galaxy as barred if: (1)~the $e$ rises smoothly to a global maximum,
$e_{\rm bar} > 0.25$, while the PA remains relatively constant (within 
$20^{\circ}$), 
and (2)~the $e$ then drops by at least 0.1 and the PA 
changes at the transition between the bar and disk region. These criteria 
have been shown to work well in identifying barred galaxies (e.g., Knapen 
et al. 2000; Jogee et al. 2002a,b, 2004; Laine et al. 2002).
An example of the  overlays and radial profiles of a barred cluster galaxy are shown
 in Figure~\ref{overlay}.  The semi-major axis of the bar $a_{\rm
   bar}$ is taken as the radius
where the ellipticity reaches a global maximum ($e_{\rm bar}$). The
disk semi-major axis length $a_{\rm disk}$ and ellipticity $e_{\rm
  disk}$ are measured from the radius of the last fitted
isophote.

The moderately inclined galaxies, which do not satisfy the bar criteria are 
classified as unbarred. This category includes clearly unbarred cases, as well as 
cases, which we denote as  `PA twist'. The latter are systems  where all the 
bar criteria are satisfied, except for the criterion of constant 
PA in the bar region:  rather than being constant within $20^{\circ}$, 
the PA of the high ellipticity feature may twist slightly more than this limit.
Some of these `PA twist' systems may actually be barred galaxies where  the 
effects of dust and SF can cause the PA to vary more than it would in a near-IR image.
This effect is more likely to happen with weak bars, where the dust lanes along the 
leading edges of the bar are curved, producing a `twisting' in the PA radial profile 
(Athanassoula 1992b). Such weak bars are also associated with SF along their leading
edge.  Among the unbarred galaxies,  we have 36 cases  of 'PA twist'. 
We use this number as an estimate of the number of barred galaxies we might be 
classifying as unbarred in the optical images, and fold it into the error bar  (upper
limit) for the optical bar fraction.

Another effect we have to address  is whether we can detect bars in the smaller/fainter
disk systems. We consider in the following analysis only disk galaxies, which we define as
galaxies with an outer disk component (e.g., S0-Sm) that may or may not be accompanied by a
central bulge (see $\S$~\ref{diskselmeth}). 
As discussed in  $\S$~\ref{datsamp}, our resolution is $\sim$~280~pc. With 
ellipse-fitting, at least 2.5 PSF elements are necessary to detect a bar. This means 
that  the lower limit on the bar radius ($a_{\rm bar}$) that we can reliably 
detect is $\sim$~700~pc. It was already noted by Kormendy (1979) that 
the sizes of bars correlate with galaxy luminosity, and late-type, fainter galaxies
host smaller bars. Erwin (2004; 2005) found that primary bars in galaxies later 
than Sbc can have radius $a_{\rm bar}$ as small as 500~pc. Thus, in order to avoid
missing small bars in late-type, faint galaxies, we make a cut 
in galaxy semi-major axis $a_{\rm disk} = 3$~kpc in addition to our magnitude cut of $M_{\rm V} = -18$.
We choose the value of  $a_{\rm disk} = 3$~kpc as a conservative cut, according
to the following analysis. 
In our cluster sample, for visually identified disk galaxies ($\S$~\ref{diskselmeth}), we find that 
$a_{\rm disk}$ and $R_{25}$ correlate with a mean 
ratio of $R_{25}$/$a_{\rm disk} = 0.87$ (Fig.~\ref{sizes}a), where $R_{25}$ is calculated 
for the cluster sample from the absolute $M_{\rm B}$ magnitude according to 
\begin{equation}
log(\frac{R_{25}}{kpc}) = -0.249 \times M_{\rm B} -  4.00,
\end{equation}
from Schneider (2006). 
In MJ07 we find that most bars have ratios $a_{\rm bar}$/$R_{25} = $0.2--0.4, where 
$R_{25}$ is the isophotal radius at which the $B$-band surface brightness reaches 25~mag~arcsec$^{-2}$. 
Assuming a median  $a_{\rm bar}$/$R_{25}\sim$~0.3, it follows that in order to ensure 
that we are looking at galaxies that host bars larger than 700~pc, we need to select galaxies 
with $R_{25} \times 0.3\sim 700$~pc, or $R_{25}\sim 2300$~pc.
Using  the mean $R_{25}$/$a_{\rm disk} = 0.87$ for our sample, this yields a 
galaxy semi-major axis $a_{\rm disk}$ of $\sim$~2.7~kpc. There 
are only 10 disk galaxies that are eliminated by this cut. Figures ~\ref{sizes}b and ~\ref{sizes}c
show the correlation of $a_{\rm bar}$ and $a_{\rm disk}$ with $M_{\rm V}$, showing only galaxies 
visually classified as disks. The dashed line
in panel ~\ref{sizes}b at 0.7~kpc represents the limit in $a_{\rm bar}$ at which we can reliably identify all bars. 
The dashed line in panel ~\ref{sizes}c at 3~kpc indicates the cut in $a_{\rm disk}$. 

In addition to quantitatively identifying and characterizing bars
using ellipse fitting, we also visually classify all galaxies in the
sample. The identification of bars through visual inspection provides
an independent check for the detection of bars through
ellipse-fits. The visual bar classification agrees with the ellipse fits for 
over 90\% of cases. For the cases where a bar is found through visual classification, 
but not through ellipse-fitting, it is because dust and gas mask the bar signature, making the 
PA twist. We conservatively take the upper error bar in the optical bar fraction
as the sum in quadrature of the binomial term and the error of $+10\%$
caused by isophotal twists. Note that the error from missed bars due to isophotal twisting 
can only make the bar fraction higher.  
Representative barred galaxies from the cluster sample
are shown in Figure~\ref{bars}.

\subsection{Visual Classification of Secondary Morphological Parameters}\label{vcmeth}

For our cluster sample, we visually classify secondary morphological 
parameters such as the prominence of the bulge and the presence of gas and 
dust.  

Since we are only interested in studying large-scale bars that extend
well beyond the bulge region of the galaxy, the prominence of the
bulge is not key for determining the bar fraction. It is interesting,
however, for studying and interpreting correlations between bar and host disk
properties (see $\S$~\ref{fbarBT}--\ref{fbarcolor}). Our goal is not to finely measure the bulge-to-total light
($B/T$) ratio in galaxies, but to identify galaxies with extreme $B/T$, such as 
systems that appear nearly bulgeless and likely have very low $B/T$,
and those with prominent bulges, suggestive of high $B/T$.  We thus 
classify galaxies into three broad groups: `pure bulge' (Fig.~\ref{morphex}a,b),
`pure disk' (Fig.~\ref{morphex}g--j), and 
`bulge+disk' (Fig.~\ref{morphex}c--f).  `Pure disk'
galaxies are those where no central spheroidal component is seen. 
Conversely, a galaxy is classified as a `pure bulge' if its morphology
is spheroidal and there
is no break in the brightness profile, indicative of the transition
between the bulge-dominated and disk-dominated region. 
 In addition, `pure bulge'
galaxies do not exhibit disk features such as spiral arms or stellar
bars. In our cluster sample of bright galaxies, we find that 23\%$\pm$12\% of galaxies are
visually classified as `pure disk', 60\%$\pm$10\% are classified
as 
`bulge+disk', and 17\%$\pm$1\% were
classified as `pure bulge.' 
 The values quoted are from the classifications of 
 I.M.  and the percent errors indicate the
sum in quadrature of the dispersion between classifiers and the
binomial term of the statistical error.
 The disagreement is due to the inherent difficulty in
separating ellipticals from disk galaxies, when the disk is smooth and 
has no unambiguous disk signature, such as a bar or a spiral arm. 

Seven members of the STAGES team performed an independent
visual classification of the sample using the standard Hubble Type system. The 
agreement between our classifications and theirs on whether a particular 
galaxy is a disk galaxy was 70\%. If their sample of visually selected disks 
is used for the analysis in $\S$~\ref{results}, our results on the optical bar
fraction do not change. Note that the standard Hubble type system is 
not optimal for our study. Principally, this is because 
Hubble types assume a correlation between the prominence of the 
bulge and the smoothness of the galaxy disk/spiral arms. While 
this correlation holds fairly well for field galaxies, it can break 
down in clusters, where there can be galaxies with large
bulge-to-disk ratios but fairly smooth disks (Koopmann \& Kenney 1998). We discuss this in more detail below. 

  In Table~\ref{morphdens_tab}, we show the breakdown of
morphological classes as a function of projected distance to the nearest cluster center for 
galaxies with $M_{\rm V} \le -18$. 
We take the core radius to be at 0.25~Mpc, because the number
density of galaxies shows a sharp break at this radius 
(Heiderman et al. 2008). The outer region is defined as lying between
the core radius at R~=~0.25~Mpc and the virial radius of the cluster,
R$_{\rm vir} = 1.2$~Mpc  (Heymans et al. 2008).  
Beyond the virial radius is the outskirt region. 

We also visually classify galaxies into those with a clumpy or smooth 
disk, motivated by the following considerations.  
Firstly, the presence of gas, dust, and star formation  along the
bar can prevent its detection in optical images, particularly for weak 
bars.  In weak bars, the dust lanes are curved because of weaker
shocks (Athanassoula 1992). In addition,  weaker shocks  can induce star formation
along the bar, while strong shocks are accompanied by straight dust
lanes and tend to suppress star formation along the bar (e.g.,
Elmegreen 1979; Das \& Jog 1995; Laine et al. 1999; Jogee, Scoville, \& Kenney 2005). 
The curved dust lanes and star formation regions in weak bars produce a pattern that 
causes fitted ellipses to have varying PA along the bar, and to sometimes  fail 
to satisfy the criterion of a flat PA plateau along the bar ($\S$~\ref{barchar}).
In very gas/dust-rich galaxies, even strong bars can be masked by
dust and star formation. These effects make it more difficult to
identify bars at optical wavelengths (e.g., Block et
al. 1994). Several studies (Eskridge et al. 2000; Laurikainen et
al. 2004; MJ07) show that, because of obscuration by
gas, dust, and star formation regions in the optical, the bar fraction is higher in the
infrared (IR) band by a factor of $\sim$~1.3 for galaxies at $z\sim$~0. 
In cluster environments, the correction factor for bar obscuration 
is unknown.

Secondly, it is useful to explore the relationship between clumpiness,
the visual prominence of the bulge, and bars in 
cluster environments, where the situation might well differ from 
the field. In the field, along the traditional Hubble Sequence, on
average the visual prominence of bulge and the tightness of the spiral
arms increase from Sd to Sa, while the clumpiness of the spiral arms decreases.  
In field galaxies, there is a wide range of $B/T$ for each Hubble type, with 
low $B/T$ galaxies being present across S0 to Sc  (Laurikainen et al. 2007;
Weinzirl et al. 2009; Graham \& Worley 2008), 
but  the average $B/T$  tends to fall in later Hubble types (Laurikainen 2007; 
Weinzirl et al. 2009; Graham \& Worley 2008).  
In clusters, where a number of processes, such as  ram-pressure
stripping or galaxy harassment can alter the 
gas content of galaxies, the relationship  between  $B/T$  and  gas/SF 
content or clumpiness of the disk may break down. For example, in the
Virgo cluster, the central concentration of galaxies does not
correlate with their star formation properties, as it does in the
field (Koopmann \& Kenney 1998). Wolf et al. (2008) discuss the effect
of these issues with respect to the Hubble type classifications 
performed by the STAGES team. 

Motivated by these considerations, we attempt to visually characterize the
presence of gas and dust in galaxies.
 The degree of `clumpiness' in a galaxy is used as a rough proxy
for estimation of the presence of gas and dust. We allocate galaxies into
two broad classes: (1)~`smooth' galaxies that show no patchy
obscuration by gas and dust or (2)~`clumpy' galaxies that have a lot 
of patchiness indicative of the presence of gas and dust. 
We find that 73\%$\pm$2\% (551/750) of the bright galaxies
in our supercluster sample appear mostly smooth (contain little or no gas and
dust), while 27\%$\pm$2\% (199/750) of the bright galaxies
appear clumpy (contain some gas and dust). The fractions quoted are from the 
classifications of I.M. and the percent errors indicate the
sum in quadrature of the dispersion between classifiers and the
binomial term of the statistical error.
Examples of `smooth' galaxies are shown in Fig.~\ref{morphex}, panels a--d and
i--j.  `Clumpy' galaxies are shown in panels e--h of Figure~\ref{morphex}.

\section{Results}\label{results}

\subsection{Selection of Disk Galaxies in Clusters}\label{cldisks}

How well do the S{\'e}rsic and blue-cloud cut methods pick out disk galaxies
when compared to visual classification?  Out of the 762 ellipse-fitted
galaxies, 608 are visually classified as disks. This number is reduced
to 573 if only galaxies with $a_{\rm disk} > 3$~kpc are
considered.

Figure~\ref{colsers} compares the disk galaxies identified through the
three different methods: visual classification, blue-cloud color cut, and a
S{\'e}rsic cut. In this paper, the color cut is made using $U-V$ color. 
Panel (a) shows where the visually-identified disk galaxies 
lie in the rest-frame $U-V$ vs. $M_{\rm V}$ plane. Moderately-inclined,
$i < 60^{\circ}$, barred galaxies are
shown as green points, where the bars are identified through
ellipse-fitting. Even though we did not identify bars with
ellipse-fits for highly inclined galaxies  with $i > 60^{\circ}$, and
do not consider inclined systems in the rest of the study, bars
were noted in such systems during the visual classification (cyan
points). Unbarred galaxies with visually-identified spiral arms (all
inclinations) are shown in pink. The black points show galaxies
identified as disks with visual classification for all inclinations, but without a bar or
spiral arms.   The solid
line separates the red sample from the blue cloud galaxies, using 
the equation
\begin{equation}
U-V = (1.48 - 0.4 \times 0.165 - 0.08 \times (M_{\rm V} + 20.0))-0.25,
\end{equation}
derived for the STAGES sample by Wolf, Gray, \& Meisenheimer (2005), 
where $M_{\rm V}$ is the $V$ absolute magnitude 
and $U-V$ is the rest-frame color.  
Panel (b) shows where visually identified disk galaxies lie in 
the $U-V$ color vs. S{\'e}rsic index $n$ plane. Symbols are the same as in panel~(a).  
The solid line shows the cutoff of $n=2.5$, which is supposed to separate disk galaxies and
spheroids.

The technique of identifying disk galaxies as those with a S{\'e}rsic
index $n<2.5$ (Fig.~\ref{colsers}b)  picks out $69\% \pm 2\%$ (396/573) of 
galaxies visually selected as
disks. The error bars represent the statistical error. The S{\'e}rsic cut method
will pick up many of the red disks that the color cut misses, however the
S{\'e}rsic cut method might miss some early-type disk galaxies with very
prominent bulges or very clumpy galaxies with bright star formation
regions in their outer disks. In addition, the presence of an AGN will
drive the S{\'e}rsic index to high values. Figure~\ref{miss}a shows examples of 
visually-identified disk galaxies missed by the S{\'e}rsic cut. 

Our analysis suggests that the S{\'e}rsic cut misses $31\% \pm 2\%$ of
visually-identified disks. How robust is this number? We consider 
the possibility that some galaxies visually 
classified as disk galaxies  (`pure disk' or `bulge+disk') may in fact be misclassified 
ellipticals. This is most likely to happen when the disk is smooth and 
has no unambiguous disk signature, such as a bar or a spiral arm. 
As stated in $\S$~\ref{vcmeth}, it is difficult to separate a `pure bulge'
galaxy from an unbarred, smooth `bulge+disk' (e.g., S0) without spiral
arms. In addition, unbarred `pure disk' galaxies without spiral arms
that appear mostly smooth could also be misclassified ellipticals. 
As a firm lower limit to the number of visually-identified disk galaxies missed by 
the S{\'e}rsic cut we consider disk galaxies (`pure disk' or `bulge+disk') that have a
clear disk signature such as a bar and/or spiral arms. This sets a firm lower
limit on the number of disk galaxies that are missed by a S{\'e}rsic cut.
We find that at least 25\% (67/267) of the
galaxies with $n > 2.5$ display unambiguous disk signatures. Thus, in
summary, we estimate that 25\% to 31\% of visually-identified disk
galaxies (e.g., S0-Sm) are missed by taking a S{\'e}rsic cut ($n < 2.5$).

The technique of selecting blue cloud galaxies picks out
out $49\% \pm 2\%$ (279/573) of
the visually-identified disk galaxies. The 294 galaxies missed are in the red sample
and the large number of these galaxies is consistent with the high number of red disks
in a cluster environment. Figure~\ref{miss}b shows some examples of visually-identified
disk galaxies in the red sample, which would be missed if a blue-cloud color cut is used
to pick out disks. 

It is interesting to look at the composition of
the red sample in more detail. The 294 visually-identified disks make
up 75\% of the total population of 390 red sample 
galaxies. Galaxies classified as `pure bulge' (e.g., E's) make up 25\%
(96/390).  Out of
the galaxies visually identified as disks in the red sample, 95\%
(279/294) are classified as `bulge+disk'  and only 5\% (15/279) are
classified as `pure disk' with no visible bulge component. 

The large proportion of the red sample consisting of visually
identified disk galaxies (e.g., S0-Sm) may seem surprising if one typically thinks
of the red sample as made up mostly of ellipticals. However, 
Wolf et al. (2008)
have shown that the red sample in the cluster contains both galaxies on the 
red sequence and also dusty red disk galaxies which do not lie on the sequence. Again, we
set a firm lower limit to the disk galaxies in the red sample, 
by considering disk galaxies (`pure disk' or `bulge+disk') that have a
clear disk signature such as a bar and/or spiral arms.
This gives a robust lower limit of 22\%  (84/390) of galaxies in the
red sample that are disks.  Thus, in
summary, our results suggest that 22\% to 75\% of the red sample is
made up of disks, with the large range primarily caused by the
difficulty in differentiating red, featureless S0-type galaxies from
spheroidals (see Figure~\ref{morphex}). A significant fraction of dusty, red disk galaxies in the
supercluster sample is also found by
 Wolf et al. (2008), where the properties of these galaxies are discussed in detail. 

In summary, we have explored how three commonly used methods for selecting 
disk  galaxies  in the field, namely,  visual classification, a single component 
S{\'e}rsic cut ($n \le 2.5$),  and  a blue-cloud cut, fare in  the A901/902 cluster  environment.
We found  that  the  S{\'e}rsic cut and blue-cloud cut methods suffer from  serious limitations, 
and  miss 31\% and 51\% , respectively, of visually-identified  disks, 
particularly the many red, bulge-dominated disk galaxies in clusters.
In cluster  environments, the   latter two methods are not well suited to 
reliably picking disk galaxies. Thus, unless otherwise stated, we use the visual 
classifications to define a  disk galaxy sample in the remaining analysis.

\subsection{Global Optical Bar Fraction}\label{optfbar}

The optical fraction of barred galaxies \textit{among all galaxies} brighter
than $M_{\rm V} = -18$, is $25\%^{+10\%}_{-2\%}$. 
However, this number is not very useful as changes in this
number can reflect a change in the disk fraction, as well as the
fraction of disks that host bars.
Furthermore, stellar bars are  $m = 2$ instabilities that
occur only in disks, and insights into their formation and evolution
can be best gleaned by inspecting the fraction of disks that are
barred at different epochs and in different  environments.
 
As mentioned in $\S$~\ref{diskselmeth}, this has motivated the 
definition of the bar fraction as 
the fraction of disks that are barred as given by Equation 1.
All studies of bars to date (e.g., deVaucouleurs 1963; Sell wood \&
Wilkinson 1993; Eskridge et al. 2000; Knapen et al. 2000; Mulchaey \&
Regan 1997; Jogee et al. 2004; Laurikainen et al. 2004; Elmegreen et al. 2004; Zheng et
al. 2005; Buta et al. 2005; MJ07;
Men{\'e}ndez-Delmestre et al. 2007; BJM08; Sheth et al. 2008) have
adopted  this definition and thus provide complementary
comparison points for our studies.

For the STAGES cluster,  we use visual classification  to define 
a disk galaxy sample (see $\S$~\ref{diskselmeth} and $\S$~\ref{cldisks}) 
and calculate the optical bar fraction  $f_{\rm bar-opt}$.
We find $f_{\rm bar-opt} =$~34\%$^{+10\%}_{-3\%}$. 
This value is similar to the optical bar fraction $f_{\rm bar-opt-EDisCS}\sim$25\% found  
for galaxies brighter than $M_{\rm V} = -19$
in intermediate-redshift ($z\sim$~0.4--1.0) clusters by Barazza et
al. (2009). 

For completeness, we also calculate the bar fraction using a blue-cloud color cut and
S{\'e}rsic cut to select disk galaxies. The results are shown in Table~\ref{fbarMV_tab}
for bright ($M_{\rm V} \le -18$) galaxies, and in Table~\ref{fbarM_tab} for galaxies with
M$_{*}$/M$_{\odot} > 10^{9}$.  Although the three disk selection methods pick 
very different number of disks  (Tables~\ref{fbarMV_tab}, \ref{fbarM_tab}, and $\S$~\ref{cldisks}), they 
yield a similar optical bar fraction $f_{\rm bar-opt}$  in the range of 29--34\%,  
This result means  that the optical bar fraction in blue
galaxies picked out by the color cut and that in low S{\'e}rsic index
galaxies, is similar to the total average bar fraction found through
selecting disk galaxies by visual classification (Tables~\ref{fbarMV_tab}  and \ref{fbarM_tab}).

\subsection{Optical Bar Fraction as a Function of the Prominence of the Bulge}\label{fbarBT}

We explore the relationship between the optical bar fraction and host galaxy properties,
such as the prominence of the bulge.

While we did not perform a structural bulge+disk+bar decomposition to 
accurately characterize $B/T$ (e.g., Laurikainen et al. 2007, Weinzirl et al. 2009), we can use the
three broad visually-classified groups of galaxies: `bulge+disk', `pure disk', and 
`pure bulge'.  

We plot the optical fraction of bars as a function of
morphological class in Figure~\ref{bfBTre}a. Here the morphological classes have
been grouped by the visual prominence of the bulge. Galaxies with a `bulge+disk'
component are in the first bin, while `pure disk' galaxies are in the
second bin. We find that $f_{\rm bar-opt}$
increases from 29\%$^{+10\%}_{-3\%}$ in `B+D' galaxies to 49\%$^{+12\%}_{-6\%}$ in
`pure disk' galaxies, suggesting that  the optical bar fraction rises 
in spiral galaxies, which are disk-dominated and have very low  bulge-to-disk 
ratios. This result is also shown in Table~\ref{fbarmorph_tab}

This result is further suggested by Figure~\ref{bfBTre}b, which
shows the optical bar fraction as a function of central concentration in the
host galaxy, as characterized by the effective radius normalized to
the disk radius, $r_{\rm e}$/$a_{disk}$. The effective radius $r_{\rm
  e}$ is calculated from single-component S{\'e}rsic fits (Gray et
al. 2008). The disk semi-major axis $a_{disk}$ comes from the
semi-major axis of the outermost ellipse fitted to each galaxy, where
the isophotes reach sky level (see
$\S$~\ref{barchar}).  The optical bar fraction clearly
increases with decreasing central concentration, from 15\%$^{+11\%}_{-4\%}$ in galaxies with 
high concentration ($r_{\rm e}$/$a_{disk}$=0.15), to 50\%$^{+13\%}_{-9\%}$ in galaxies 
with low  concentration ($r_{\rm e}$/$a_{disk}$=0.75). 
We note that the optical bar fraction does not show a similar 
correlation with S{\'e}rsic index $n$ for this sample, as
there is a large scatter of $n$ within each morphological class. 
In addition, the relationship of the optical bar fraction with $r_{\rm e}$/$a_{disk}$
in Figure~\ref{bfBTre}b can be compared with previous studies (BJM08),
which show a similar trend in the field.

At this point, it is important to ask whether the trend of a higher $f_{\rm bar-opt}$  
in disk  galaxies with no significant bulge component is real, or due to 
systematic effects  which cause us to miss primary bars in galaxies with prominent 
bulges. In this paper, we are considering large-scale primary bars, which by 
definition lie outside the bulge region. If the bar is strong and/or extended
well beyond the bulge region, it is unlikely that the ellipse-fit method and 
quantitative criteria described in $\S$~\ref{barchar} would miss the bar.
On the other hand, if the bar is  only slightly larger than the bulge, one may face
cases where  the ellipse-fit method might miss the bar.  
Furthermore, if the bar is intrinsically weak (i.e. of low ellipticity), then the 
dilution effect of the large bulge may cause the measured ellipticity of the
weak bar  to  fall below  the cutoff value (0.25) where it would be considered
a bar. We tested and assessed these effects in several ways described below.

Firstly, we note that studies using a method different from ellipse fitting, namely 
2D bulge-disk-bar decomposition on nearby galaxies, also show that galaxies with
a larger bulge-to-total ratio  ($B/T$) host a lower proportion of bars than  
galaxies of lower $B/T$  (Weinzirl et al. 2009). For disk galaxies with
 $M_{\rm B} < -19$, the bar fraction increases from 31\%$\pm$13\% in spirals 
with $B/T \ge 0.4$, to 68\%$\pm$4\% in spirals with $B/T \le 0.2$ (see their 
Table 8 and $\S$~5.6).  Of course, one could argue 
that  2D bulge-disk-bar decomposition is also more likely to miss the bar 
component, when the bar contains a much smaller fraction of the light than
the bulge.  
We therefore performed a second test. In addition to using ellipse-fitting to 
detect bars, we also visually inspect all galaxies as an extra check. We 
expect that we would see most short bars  in galaxies with large bulges  
via visual classification. We only find two such cases. This 
small number is not enough to make up for the large drop in  $f_{\rm bar-opt}$ 
toward bulge-dominated galaxies.
 
We also performed a third test. If the trend of a lower $f_{\rm bar-opt}$ 
in  bulge-dominated galaxies was due to the fact that  a prominent bulge 
causes us to systematically miss bars with low ellipticity around 0.25, 
then we would not expect the trend to persist if we only include strong
(high ellipticity) bars.  We tested this by  recomputing the optical bar 
fraction  in pure disks and B+D systems after applying a lower limit 
cutoff  0.4 and 0.5 on  the  bar  ellipticity. In both cases, we find that
the trend of a  higher $f_{\rm bar-opt}$  in disk galaxies without prominent
bulges remains. We conclude that the latter trend is likely real, and will explore theoretical
scenarios that could account for it in $\S$~\ref{disc}

We also note that the rise in the optical bar fraction as a function of the 
prominence of the bulge or central concentration of the host galaxy is in
agreement with BJM08, which found that the optical bar fraction in
pure disk galaxies is a factor of $\sim$~2 higher than in  disk galaxies with 
prominent bulges, for an SDSS sample of $M_{\rm V} \le -18.6$
blue-cloud galaxies and redshift range 
$0.01 \ge z \ge 0.03$. This result is confirmed by A09, 
who find that the optical bar fraction increases
from 29\% in S0 galaxies, to 54\% in late-type (Sc-Sd) systems, using
SDSS galaxies at  $0.01 \le z \le 0.04$.

\subsection{Optical Bar Fraction as a Function of Host Luminosity}\label{fbarL}

In Figure~\ref{fbarlum}a-c we show the optical bar fraction as a function of host galaxy
rest-frame  magnitude  $M_{\rm V}$. The optical bar fraction is calculated for
all three methods of disk selection (color cut, S{\'e}rsic cut, and visual
classification). For all three methods of disk selection, the optical bar fraction shows a decrease
from $\sim$~60\%$^{+14\%}_{-10\%}$ at $M_{\rm V}=-21.5$ to $\sim$~20\%$^{+11\%}_{-4\%}$
at $M_{\rm V} = -18.5$. 

This result may seem counter-intuitive given the fact that we find a lower optical 
bar fraction in bulge-dominated galaxies, and we might expect  such systems to be
on average  brighter. 
However, Table~\ref{fbarMVmorph_tab} explains why we find the 
opposite result. This table shows how the optical bar fraction 
varies as a function of morphological class and absolute magnitude. Here the morphological classes 
refer to the four visually-classified disk morphological
classes: `bulge+disk smooth', `bulge+disk clumpy', `pure disk smooth', and
`pure disk clumpy'. Table~\ref{fbarMVmorph_tab} shows that the optical bar fraction is higher 
at brighter $M_{\rm V}$ \textit{for any given morphological  class}. 
Therefore, when all of the visual morphological classes are grouped together 
and $f_{\rm bar-opt}$ is calculated  as a function of $M_{\rm V}$ in Fig.~\ref{fbarlum}c,  
the optical bar  fraction is higher for brighter magnitudes. 

This result is consistent with  the findings of Barazza et al. (2009) for 
cluster galaxies at intermediate redshifts ($z\sim$~0.4--1). This study also finds that, 
although brighter, early  Hubble type galaxies host fewer bars than fainter, late-type 
galaxies, within a given Hubble type, brighter galaxies on average have a higher 
optical bar fraction. 

\subsection{Optical Bar Fraction as a Function of Host Color}\label{fbarcolor}

We find no significant difference in the optical bar fraction in disks on 
the red sequence and blue cloud. When disks are selected through visual 
classification, the  optical bar fraction on the red sequence is 
$f_{\rm bar-RS} \sim$~34\%$^{+11\%}_{-4\%}$ and on the blue cloud, it 
is $f_{\rm bar-BC} \sim$~34\%$^{+11\%}_{-4\%}$.  This can be easily seen 
by inspection of Fig.~\ref{colsers}. The identical values for the blue 
cloud and red sequence explain in part why  the global optical bar fraction 
$f_{\rm bar-opt}$  based on visual selection of disks, is similar to the
one obtained by selecting  disks via a blue cloud cut.

Taking a global average of the optical bar fraction across the 
blue cloud and red sequence may not reveal the 
dependence of the optical bar fraction solely on color 
because the relative number of 
bright to faint galaxies is different on the blue cloud and red 
sequence, with the red sequence having more bright galaxies  (Fig.~\ref{colsers}).
As already shown in $\S$~\ref{fbarL}, the bright systems have a higher 
optical  bar  fraction than fainter galaxies,
since the  optical bar fraction  rises at higher luminosities for 
each given morphological type (Table~\ref{fbarMVmorph_tab}). 
Therefore, we expand the exploration of the optical bar fraction 
by looking at the breakdown of the optical bar 
fraction as a function of rest-frame $U-V$ and  $M_{\rm V}$  (Table~\ref{fbarMVcolor_tab}), 
as well as $U-V$ and   visual morphological class  (Table~\ref{fbarcolormorph_tab}).  

In  Table~\ref{fbarMVcolor_tab}, we find that most bright galaxies 
are red with $U-V$ color $>1$ and $f_{\rm bar-opt}$ of 44\%--69\% 
at $M_{\rm V}=-20$ to $-22$. 
Table~\ref{fbarcolormorph_tab} shows that for `bulge+disk' galaxies that are red ($U-V > 1$)
galaxies classified as `clumpy' have the highest bar fraction
of 76\%. Blue pure disk galaxies
have an optical bar fraction of $\sim$~50\%. 

\subsection{Optical Bar Fraction as Function of Kappa, $\Sigma_{10}$, ICM
  density, and Distance to Nearest Cluster Center}\label{fbardens}

How does the local environment affect the optical bar fraction, and where do
barred galaxies live with respect to the density peaks in the
A901/902 cluster environment? In this section, we make a first step in 
exploring these questions
using four traces of local environment density: the line-of-sight projected
surface mass density $\kappa$ (Heymans et al. 2008), the  local galaxy
number density $\Sigma_{10}$ (Wolf, Gray, \& Meisenheimer 2005; Gilmour et al. 2007), 
the ICM density as characterized by the X-ray emission from hot intra-cluster
gas in counts, and the projected distance to the nearest cluster center. We
calculate $\Sigma_{10}$ by
finding the radius enclosing the ten nearest neighbors to a
galaxy. This is used to calculate a galaxy number density, quoted in
$(Mpc/h)^{-2}$. Maps of $\kappa$ for the Abell 901/902 field are constructed by 
Heymans et al. (2008) through an analysis of weak gravitational 
lensing, which is sensitive to the line-of-sight projected surface mass 
density.  

Figure \ref{stuffvdmin} shows the variation of the three measures of local environment
density ($\kappa$, $\Sigma_{10}$, and ICM density) with distance to
the nearest cluster center. It is evident from all three tracers, 
that local density decreases with increasing distance from the nearest cluster
center. The core, outer region, and outskirt of the clusters are defined 
in $\S$~\ref{vcmeth}. 
One caveat in this analysis is that the quantities used are projected 
quantities. 

Figure \ref{fbdens} shows the variation of the optical bar
fraction function of: (a)~distance from nearest cluster
center, (b)~log~$\Sigma_{10}$, (c)~$\kappa$, and (d)~ICM
density. 
We find that between the core and the virial radius of the cluster 
($R\sim$~0.25 to 1.2 Mpc), the optical bar fraction $f_{\rm bar-opt}$ 
does not depend strongly on the local environment density tracers  ($\kappa$, 
$\Sigma_{10}$, and ICM density), and varies at most by a factor of $\sim$~1.3, 
allowed by the error bars.

Within the core region, the small  number statistics and projection effects make it hard to draw 
a robust conclusion on the detailed variation of the optical bar fraction.
In fact, the detailed behavior seen as we move from the outer region to  the 
cluster core varies according to which indicator  is used:  $f_{\rm bar-opt}$ shows 
no change when using projected radius (Fig.~\ref{fbdens}a), dips by a 
factor of $\le$~1.5 when using $\Sigma_{10}$ (Fig.~\ref{fbdens}b), or 
$\kappa$ (Fig.~\ref{fbdens}c), and rises by a factor of  $\le$~1.2 
when using the  ICM density (Fig.~\ref{fbdens}d). 
Given the small number statistics, projection effects, and the fact that different
indicators suggest different trends in the cluster core, we can only
say that inside the cluster core, 
we do not find  evidence for a variation stronger  than a factor of 1.5  
in the optical bar fraction $f_{\rm bar-opt}$ as a function of any 
of the three environmental indicators in Fig.~\ref{fbdens}.

How do our results compare to other studies?
The recent study of  bars in field and intermediate density regions  
by A09 reports no variation of the 
optical bar fraction with $\Sigma_{5}$, where $\Sigma_{5}$ varies
between -3 and 2.  
On the other hand,  several previous studies have found an 
enhanced optical bar fraction toward cluster centers 
(Barazza et  al. 2009; Thompson 1981; Andersen 1996).
We discuss the implication of the results from our study as 
as well as these other works in $\S$~\ref{disc}.

\subsection{Comparison of the Optical Bar Fraction in the A901/902 Clusters and 
the Field}\label{fieldcomp}
 
To further explore the  impact of environment  on the evolution of
bars and disk galaxies, it would be desirable to compare the properties 
of disk galaxies in cluster and field  samples, which are at similar 
redshifts and are analyzed in a similar way.
We do not have a field sample at the same redshift as that of the 
A901/902 supercluster ($z\sim$~0.165),  and therefore resort to an 
approximate comparison only, bearing in mind the caveats. 

We compare the results on bars and disks from the STAGES sample to those 
from studies of nearby galaxies by   MJ07 and A09. 
In these  studies, bars are identified and characterized through 
ellipse-fits, as for our STAGES study. 
The sample of  MJ07 is based on moderately inclined galaxies in 
the Ohio State University Bright Spiral Galaxy Survey (OSUBSGS;
Eskridge et al. 2002), which contains galaxies of RC3 type S0/a or later, 
(0$\leq$ T $\leq$9), $M_{B}< 12$, $
D_{25}< 6\arcmin.5$, and $-80^\circ <\delta< +50^\circ$.
This sample is dominated by early to intermediate-type (Sab-Sc) galaxies, 
in the range $M_{\rm V} = -20$ to $-22$. The galaxies are local field 
spirals, and strongly interacting galaxies are not included in the MJ07 analysis.

The sample of  A09 is based on the  Sloan Digital 
Sky Survey (SDSS; Abazajian et al. 2004) within the redshift 
range $z\sim$~0.01--0.04 and with $M_{\rm r} > -20$. 
Hubble types for 
galaxies in this sample are from visual classification, bars are identified
through ellipse-fitting and environment
density is estimated from the projected local galaxy density ($\Sigma_{5}$). 
The densities considered
in the A09 sample range from the field to intermediate densities comparable to 
those in the outer regions and outskirts
of our clusters ($\Sigma_{5}$ from -3 to 2).

Before we compare the results obtained for bars in the cluster with those from
the field studies, we compare the properties (e.g., absolute magnitude, color)  
of the underlying galaxy populations  in the field and cluster samples.
In Figure~\ref{sampcolMV}a--b, we compare the distributions
of $M_{\rm V}$ absolute magnitude and rest-frame $U-V$ color
for the STAGES cluster, OSUBSGS, and SDSS field samples. The SDSS data are from A09. 
The OSU sample
is brighter, and is dominated by galaxies in the range 
$M_{\rm V} = -20$ to $-22$, while the STAGES sample is dominated by
galaxies in the range $M_{\rm V} = -18$ to $-20$. The SDSS sample spans a narrow range in 
$M_{\rm V} = -19.5$ to $-22$.  The STAGES and OSU samples have 
a similar range in $U-V$ color, although the OSU sample has a slightly higher 
proportion of bluer galaxies.

We have shown that the optical bar fraction is a strong function of
galaxy morphology and luminosity. It is therefore important to compare not only the
global optical bar fraction, averaged over all galaxy types for a given
magnitude cut, but also to compare galaxies of different morphological types,
namely spiral galaxies with prominent bulges and spiral galaxies that appear 
as pure bulgeless disks.  The latter  galaxies are also of  particular interest
as they present a potential challenge to hierarchical $\Lambda$CDM models.

Table~\ref{fbarfield_tab} shows the detailed comparison of the optical bar fraction  between  
field and cluster galaxies. The comparison  is done separately for 
very bright (parts A and B of Table~\ref{fbarfield_tab}) and moderately bright to faint 
(part C of Table~\ref{fbarfield_tab}) galaxies. The global optical bar fraction 
averaged over all galaxy types in the samples, as well as the optical bar fraction 
for galaxies classified as  `bulge+disk' (B+D), and galaxies with pure disks, are shown.
 
The upper error bar on the optical bar fraction quoted for this study 
and MJ07, is the sum in quadrature of the error in the bar fraction from 
isophotal twists ($\S$~\ref{barchar}) and the statistical error. 
Note that including isophotal twists into the optical bar fraction can 
only make the optical bar fraction higher. Therefore, the lower error bars 
quoted represent only the  statistical error. 

When comparing bright galaxies in STAGES with  the OSU field survey (part A in Table~\ref{fbarfield_tab}),  
we find that the average global optical bar fraction, as well as the optical bar fraction 
for galaxies with B+D  is slightly higher. However,  the difference is not 
significant within the error margins. 

When comparing bright  galaxies in STAGES with the SDSS field survey (part B in Table~\ref{fbarfield_tab}),  
we find that $f_{\rm bar-opt}$ for the STAGES cluster sample  is higher
than the field by a factor of $\sim$~1.2. In Table~\ref{fbarfield_tab} part C, we show the comparison of 
the optical bar fraction for faint galaxies $M_{\rm V} = -18.6$ to $-20.5$ 
between the STAGES cluster sample and the SDSS sample. We find that the optical bar fraction 
for early-type `B+D' galaxies in the cluster is lower by a factor of 0.6. For late-type `pure
disk' galaxies, the optical bar fraction in the cluster is higher by a factor 
of 1.2. 

In summary, for bright early Hubble types, as well as faint late-type
systems with no evident bulge, the optical bar fraction in the Abell 901/2  
clusters is comparable within a factor of 1.2  to that of  field galaxies 
at lower redshifts ($z < 0.04$).

\subsection{Bar Strength Distribution in the A901/902 Clusters and 
the Field}\label{fieldae}

Since we have found that the optical 
bar fraction is a strong function of $M_{\rm V}$, for the following 
analysis, we focus on galaxies brighter than $M_{\rm V} = -20$ in 
all samples.
Figure~\ref{ebosu}a--b shows the peak ellipticity
$e_{\rm bar}$  distributions for the STAGES and OSUBSGS
samples, respectively. In panel (a) the pink and green lines show the
$e_{\rm bar}$ distributions for galaxies classified as `bulge+disk'
and 'pure disk', respectively. In panel (b) the distributions are
split into bulge-dominated galaxies (S0-Sbc; pink) and (Sc-Sm;
green). 

We find that in both the cluster and field, the highest
ellipticity bars lie in disk-dominated galaxies, although number statistics
for this group are low in both samples, for galaxies brighter than $M_{\rm V} = -20$. In the A901/902
cluster system at $z = 0.165$, $e_{\rm bar}$ peaks at lower values
($e_{\rm bar} \sim$~0.5) than in lower-redshift field OSUBSGS galaxies ($e_{\rm bar} \sim$~0.7). 

The result of lower $e_{\rm bar}$ in STAGES compared to 
OSU could be caused by more bulge-dominated hosts in STAGES
than OSU.  Bars in galaxies with
large bulges can appear weaker (i.e., rounder). This effect has
been  observed in the STAGES sample, as well as in SDSS by BJMO8, and could be an 
artifact due to the apparent dilution of the ellipticity of the bar isophotes by the bulge. 
If it is not an artifact, it is possible that a massive bulge can
affect the actual bar supporting orbits and cause the bar to become rounder. 
 The same result is seen using $Q_{\rm g}$ (the  maximum gravitational torque induced by the bar normalized
to the axisymmetric component)  for characterizing bar strength 
(Laurikainen et al. 2007).

However when measuring the Fourier amplitude
of the bar to characterize bar strength, early-type galaxies appear
to host stronger bars (e.g., Laurikainen et al. 2007). The bar-to-total 
mass ratio also increases toward early-type galaxies, although with large
scatter, as measured from 
2D bulge-disk-bar decomposition (Weinzirl et al. 2009).

\section{Discussion}\label{disc}

We can use clusters as a laboratory for learning about 
the interplay between internal and external 
drivers of galaxy evolution. Bars are the most efficient internal
drivers of galaxy evolution, however it is still an open question what
makes one galaxy more susceptible to bar formation than another, and 
how bars evolve as a function of epoch and environment.
The situation is complex because, in principle, the fraction of barred 
galaxies in a cluster depends on the epoch of bar formation, the 
robustness of bars, the interplay between cluster environmental 
processes (harassment, tidal interactions, ram pressure stripping), and 
the evolutionary history of clusters. 
 
The relationship between the bar and the properties of its host galaxy, 
such as the Hubble type or bulge-to-total ($B/T$) ratio has been 
explored in several studies (e.g., Odewahn 1996; BJM08; A09; Weinzirl 
et al. 2009; Laurikainen et al. 2009), focusing mainly  on field galaxies.
We first discuss results reported to date on how the bar fraction varies 
from intermediate to late Hubble types (Sbc to Sd or Sm).  
The early study by Odewahn (1996) 
found that the optical fraction of strong bars rises from intermediate
to late Hubble types (e.g., from Sbc to Sm).
The study by BJM08, where the sample of disk galaxies was dominated by 
galaxies of intermediate to late Hubble types, also 
found that the optical bar fraction rises in galaxies that tend 
to be disk-dominated and devoid of a bulge. 
Similarly, in the study by Weinzirl et al. (2009), 
the near-IR bar fraction was found to be larger toward systems of low
bulge-to-total ($B/T$) ratios. 
Thus, the trend of a higher bar fraction in 
disk-dominated systems is reported by at least three studies. 
Our results in the study of A901/902 (see below) are also in agreement
with this trend.

As we go from intermediate to early Hubble types, (e.g., 
from Sbc to S0/a to S0), two recent studies seem to agree that the
bar fraction  is much lower in S0 than among  galaxies of type Sbc
to S0/a.  Laurikainen et al. (2009)  report that the NIR  bar fraction 
first rises in going from spirals (Sa-Scd) to S0/a,  and then falls sharply among S0 galaxies. 
The study by A09, based on SDSS galaxies,  finds an optical 
bar fraction of  29\% in S0,  compared  to  55\%  in early type (S0/a-Sb), 
and 54\% in intermediate-to-late type (Sbc-Sm).  
Thus, they confirm that the bar fraction drops sharply among S0s, but
unlike Laurikainen et al. (2009), they do not find evidence for an 
increase from Sbc to S0/a. 

In summary,  when considering all the studies to date  
(Odewahn 1996; BJM08; Weinzirl et al. 2009; Laurikainen et al. 2009; A09) 
it appears that 
{\it  the bar fraction is highest in late type Sd-Sm disk-dominated galaxies} 
and
{\it  lowest among S0}, while 
conflicting results exist on how the bar
fraction varies from  S0/a to Sc.

How do our results in the A901/902 supercluster compare to the above results in the 
field? We have found that the optical bar fraction $f_{\rm bar-opt}$  
in the A901/902 cluster system depends on {\it  both} 
the bulge-to-disk ratio and the luminosity ($\S$~\ref{fbarBT} and \ref{fbarL}).
We do not  have quantitative measures of bulge-to-total ($B/T$) ratios 
in order to classify galaxies into Hubble types, but  we separate
spirals into two broad classes: those with `bulge+disk' and those that are  
`pure disks'.  
We then found that  at a given luminosity,  $f_{\rm bar-opt}$ is higher 
among galaxies, which are `pure disks', without  a significant bulge 
component, as compared to those with a bulge.
In $\S$~\ref{fbarBT}, we explored whether this trend could be artificially 
caused  by systematic effects whereby  a more prominent bulge might cause 
us to systematically miss primary bars. We concluded that this was
unlikely, and that the trend of a higher  $f_{\rm bar-opt}$ in galaxies 
without a bulge, as opposed to those with bulges, is  a robust one.
This trend is in agreement with the above-described results from 
earlier studies  (e.g.,  Odewahn 1996; BJM08; Weinzirl et al. 2009).

In  addition, we have found a new hitherto unappreciated dependence 
of the bar fraction on luminosity. Specifically, we find that
for a given visual morphological class,   
$f_{\rm bar-opt}$ rises  at higher absolute magnitude.  
A concurrent study by  Barazza et al. (2009) similarly report 
that for cluster and field galaxies at  $z = 0.4 - 0.8$ with early 
Hubble types S0-Sb, the  bar fraction rises for brighter galaxies.
{\it Our results thus suggest that the relationship between bar fraction and  
bulge-to-disk ratio may not be a monotonic one, and may depend on other 
factors, such as the  gas content or luminosity.}

How do our results and those reported from other studies 
fit within theoretical scenarios of bar formation?
Let us first consider the trend of a higher $f_{\rm bar-opt}$  in galaxies
classified as `pure disk'.
In one theoretical scenario, it has been suggested that bars can form and be 
maintained through the swing amplification of gravitational
instabilities (e.g., Toomre 1981; Binney \& Tremaine 1987) in dynamically cold disks. 
The presence of a significant amount of cold gas in the disk lowers the
Toomre $Q$ parameter, favoring the onset on gravitational instabilities.
Typically $Q<1.5$ is needed for efficiently maintaining the swing amplifier.
Such bars are less likely to grow in galaxies where a prominent
bulge leads to an inner Lindblad resonance (ILR), which cuts off the
 feedback loop for swing amplification (by preventing stellar spiral 
density waves from going through the center of the galaxy).   
The existence of an ILR requires not only  the presence of a bulge, but  
requires  a large  $B/T$ in order to produce a large enough density 
contrast  between the inner and outer regions of the disk.
Therefore, in this scenario, it is expected that galaxies with a large 
gas mass fraction and/or with no ILR  are more likely to host bars 
than galaxies, which are gas-poor and have ILRs, for instance, due to
a prominent bulge.

Our result in A901/902, and the results in the field by BJM08, whereby 
the  bar fraction is higher in galaxies with pure disks than in galaxies 
with bulges, seem broadly consistent with this scenario. 
We also note that  the drop in bar fraction among S0s  (Laurikainen et
al. 2009; A09)  is also consistent with this framework, since S0s have
prominent bulges, likely host ILRs, and have a low gas mass fraction.
One would like to know whether it is the  $B/T$ ratio or the 
gas fraction that  primarily control the bar fraction.
For field galaxies  where the pure disk spirals  (Sd) have 
both a low $B/T$ and  a large gas fraction,  it is difficult to 
disentangle these two factors.  However, the fact that we see 
a higher  $f_{\rm bar-opt}$  among pure disks  in {\it clusters} 
suggests that  the gas content of 
the disk is less relevant, since cluster galaxies can be stripped of their
gas by various cluster processes (e.g., Balogh et al. 2000; Quilis et al. 2000).
For the A901/902 cluster system, we find that 73\%$\pm$2\% of galaxies 
appear smooth (i.e., contain little or no patchiness caused by dust/gas).

However,  the above scenario where bars are  formed and 
maintained through the swing amplification of gravitational instabilities 
cannot fully explain the full range of observational results to date.
For instance, our result  that  within a given morphological class, 
$f_{\rm bar-opt}$   rises at higher luminosity requires us to consider 
other theoretical aspects of bar evolution, such as the effect of the DM 
halo. 
Studies have found that the DM fraction rises for lower luminosity 
systems, although with large scatter (Persic et al. 1996; Kassin et al. 2006). 
The interplay between a DM halo and the disk  can influence both the formation
and subsequent growth of a bar.  
In early simulations with rigid DM halos, the halo
 acts as a dynamically hot component and thus  tends to make an embedded {\it unbarred}  
disk more stable against bar formation. We note that such earlier simulations (e.g., 
Ostriker \& Peebles 1973), using rigid rather than live DM halos, may have exaggerated 
the inhibition of the bar. 
In more recent simulations,  live halos are used to represent a more realistic 
view of real galaxies and bar evolution.
Debattista \& Sellwood (2000) find that in simulations with live
halos, bars form readily and are difficult to destroy. A massive live halo has the
effect of braking the bar through dynamical friction, where the amount of braking 
depends on the DM halo-to-disk mass ratio within the region of the disk. 
Athanassoula (2002, 2003) finds that the distribution of the halo mass is the most 
influential factor dictating the evolution of the bar. Bars in more
halo-dominated simulations develop more slowly than bars embedded in
disks that are massive compared to the halo in the inner
regions. However, although the bars grow more slowly, they tend to become stronger 
because the live DM halo acts a  sink for angular momentum transferred 
out by the bar. 

In the context of these simulations, one would expect bars to form and
grow more slowly in galaxies with higher DM fraction, namely in fainter galaxies. 
Our results are consistent with some 
aspects of this scenario. Our results of a higher  $f_{\rm bar-opt}$  among  brighter
galaxies may be related to the faster and more efficient growth of bars in brighter
galaxies with lower DM fraction.
However, another prediction of these simulations is that bars in brighter galaxies 
with lower DM fraction would,  in the end, be weaker. 
We do not find statistical evidence of a change  in  $e_{\rm bar}$ 
with luminosity.  The mean $e_{\rm bar} = 0.6$ for galaxies fainter than $M_{\rm V}=-20$, 
compared to the mean $e_{\rm bar}=0.5$ for galaxies brighter than $M_{\rm V}=-20$. 
This change in the  mean $e_{\rm bar}$ by  a factor of 1.2 is within the 
statistical error.

How does the frequency and evolution of bars differ in different environments?
In $\S$~\ref{fbardens}, we found that between the core and the virial radius of
the cluster ($R\sim$~0.25 to 1.2 Mpc) at intermediate densities
(log($\Sigma_{10}$)$ = 1.7$ to 2.3), the optical bar fraction $f_{\rm bar-opt}$ 
does not depend strongly on the local environment density tracers  ($\kappa$, 
$\Sigma_{10}$, and ICM density), and varies at most by a factor of $\sim$~1.3, 
allowed by the error bars. These results agree with those of A09 for
intermediate densities. A09 find no dependence of the optical bar fraction 
on local environment density, over a wide range of log($\Sigma_{5}$)$ = -2$ to 3. The 
average galaxy number density in A09 is lower than our sample, and is 
comparable to the environments present in the outer region of our 
cluster sample (Fig.~\ref{fbdens}), where 
we also find no dependence of the bar fraction with $\Sigma_{10}$. 
Recently, Romano-D{\'{\i}}az et al. (2008) used theoretical models 
to study the formation of bars in a cosmological context. Their
results  suggest that interaction with the halo sub-structure induces bars. 
Because this substructure is present in all environments, these models imply a similar
bar fraction across a large range of environment densities, which is consistent with our results.

Inside the cluster core at the highest densities, our data do not yield conclusive results for several
reasons. Firstly, the number statistics are very limited, and at best, within the 
caveats of limited number statistics, we can say that  $f_{\rm bar-opt}$ does 
not show evidence for a variation larger than a factor of 1.5  toward the
core as a function of the environmental indicators ($\S$~\ref{fbardens}). 
Secondly, the detailed behavior seen as we move from the outer region to  the 
cluster core varies according to which indicator  is used:  $f_{\rm bar-opt}$ shows 
no change when using projected radius (Fig.~\ref{fbdens}a), dips by a 
factor of $\le$~1.5 when using $\Sigma_{10}$ (Fig.~\ref{fbdens}b), or 
$\kappa$ (Fig.~\ref{fbdens}c), and rises by a factor of  $\le$~1.2 when 
using the  ICM density (Fig.~\ref{fbdens}d).  

Some early studies looked at the optical bar fraction toward the centers of 
local clusters (Thompson 1981; Andersen 1996). These studies use visual 
classification of Coma and Virgo galaxies, respectively. Both studies use the 
velocity distributions of cluster galaxies to argue that the fraction of barred 
galaxies is enhanced toward the cluster cores. Thompson (1981) finds 
that bars occur twice as often in the core compared to the outskirt region of 
the Coma cluster, while Andersen does not quote specific numbers. 
A recent study by  Barazza et al. (2009) using ellipse-fitting on 
a sample dominated by galaxies with $M_{\rm V} = -20$ to $-22$ 
at intermediate redshifts $z\sim$~0.4--1, 
finds a rise in the optical bar fraction of a factor of $\sim$~2 in cluster cores.
However they caution that this result may be affected by low number statistics. 

Although, we cannot make a conclusive statement about the behavior of
the bar fraction in the core region based on our data, we can speculate on 
what effects are at play in the cluster cores
that might affect the bar fraction and trends found in previous studies. 
The possibility raised by previous studies (Thompson 1981; Andersen 1996; 
Barazza et al. 2009) that the  optical bar fraction in the cluster core  is 
higher (or not significantly lower) 
from that in  the outer region of the cluster may at first seem puzzling 
because bulge-dominated galaxies are generally prevalent in cluster cores.  
Given our results of a lower optical bar fraction in bulge-dominated galaxies 
($\S$~\ref{fbarBT}), one might naively expect a sharp drop in the optical bar 
fraction toward the cluster cores. The fact that such a drop is not
seen, suggests that other processes in the core tend to enhance the
bar fraction, thereby countering the drop. We discuss two such
processes below. 
 
In the cluster core, the galaxy collision and interaction timescale is 
very short because of the high galaxy number density ($n$) and the  
large galaxy velocity dispersion ($\sigma_{\rm gal}$): 
\begin{equation}
t_{\rm coll} = \frac{1}{n\sigma_{\rm gal}A}, 
\end{equation}
where $A$ is the collision cross-section. 
Heiderman et al. (2008) calculate $t_{\rm coll}$ for the core, outer region, 
and outskirt of the Abell 901/902 system. 
They find that    $t_{\rm coll}$ in the cluster cores is $\sim$~0.7~Gyr, compared
to $\sim$~10~Gyr and $\sim$~200~Gyr in the outer region and outskirts, respectively.
Thus, galaxy tidal interactions are expected to be frequent in the  A901/902  
cluster cores.  This can lead to  the tidal triggering of bars in sufficiently 
cold disks and would tend to raise the optical bar fraction in cluster cores.

Another additional factor favoring bar formation  in  cluster cores is 
that the frequent tidal interactions are unlikely to develop into galaxy mergers 
or  into strong galaxy  interactions associated  with large tidal heating 
because the galaxy velocity dispersion is large (700 to 1000 km/s for the 
A901/902 system;  Gray 
et al. in prep.). The latter type of mergers or interactions tend to lead to 
strong  tidal damage and heating of the disk and could destroy the bars.
The results of  Heiderman et al. (2008) are consistent with this 
scenario:  they find that in the A901/902 clusters, the galaxy mergers and 
strongly interacting galaxies (those with strong morphological distortions) 
 are rare and tend to be located outside the cluster core, in the outer region between the 
core and virial radius. This supports the idea that the large galaxy velocity 
dispersion in cluster cores are not conducive to mergers and violent interactions.   
In effect, the core environment may well provide many frequent weak, non-destructive
tidal interactions  (harassment), which favor the triggering of bars in cold 
disks.  
In such a case, the \textit{trend of  a lower bar fraction from a population of 
galaxies with high $B/T$ in the core, may be counteracted by the opposite 
tendency for core environmental processes  (e.g., harassment) to favor 
bar formation}\footnote{One way to further test the hypothesis that tidal triggering of bars via 
harassment is important toward the core would be to look at how the 
optical bar fraction for systems with a fixed narrow range of $B/T$ 
varies within the cluster. 
Unfortunately, in the core, we do not have enough number statistics for pure 
disk  systems ($B/D\sim$~0) and no quantitative measure of $B/T$ to split the
class of `bulge+disk' systems into sub-classes with narrow ranges of $B/T$. We find
no conclusive trend of the bar fraction with density within the
clusters for the broad class of `bulge+disk' systems.}.

It is also interesting in this context to note that higher bar fractions have
been reported for binary pairs of galaxies (e.g., Elmegreen et al. 1990; Varela et al. 2004). 
Elmegreen et al. (1990) find a similar bar fraction in field and groups ($\sim$~30\%), and a 
higher fraction in  galaxy pairs ($\sim$~50\%), but only for early Hubble types. 
Their study is based on visual classifications from a number of different 
field, group, and binary samples of nearby galaxies. Binary pairs vary in 
separation, but all have projected separation distances $<180$~kpc.
Varela et al. (2004) also find that the optical bar fraction in binary pairs
is twice as high as in isolated galaxies, again only for early Hubble types. This is 
consistent with the idea that weak interactions may enhance the bar fraction.

\section{Summary and Conclusions}\label{sum}
We have used the STAGES $HST ACS$ survey of the Abell 901/902
supercluster in F606W at z$\sim$0.165 to study the properties of barred and unbarred disks
in a dense environment. Ellipse-fitting was used to identify and
characterize the properties of bars in our sample. Visual
classification was used to characterize secondary morphological
parameters such as the prominence of the bulge, clumpiness, and spiral
arms. Galaxies were
grouped into the broad classes: `pure bulge', `bulge+disk', and `pure disk'. In
addition, the galaxies were classified as either `clumpy' or
`smooth'. We find the following results:

\begin{enumerate}
\item{\textit{Disk selection in clusters:} 
To identify the
optical bar fraction $f_{\rm bar-opt}$, three common methods of disk selection were used
and compared: visual classification, $U-V$ color cut, and S{\'e}rsic
cut. We find 625, 485, and 353 disk galaxies, respectively, via visual classification, a 
S{\'e}rsic cut ($n \le 2.5$), and  a blue-cloud cut (Table~\ref{fbarMV_tab}).  
A color cut misses 51\%$\pm$2\% of visually-identified disk galaxies. 
A S{\'e}rsic cut misses 31\%$\pm$2\% of visually-identified disk 
galaxies with $n > 2.5$. Therefore, a blind application of a color cut or
 S{\'e}rsic cut would miss many of the red galaxies with prominent bulges that
are prevalent in a cluster environment.}

\item{\textit{Global optical bar fraction:} 
 For moderately inclined galaxies ($i < 60^{\circ}$), we find that
 the three methods of disk selection
(visual, color cut, S{\'e}rsic cut), we obtain a similar optical bar fraction $f_{\rm 
bar-opt}$ of 34\%$^{+10\%}_{-3\%}$, 31\%$^{+10\%}_{-3\%}$, and
 30\%$^{+10\%}_{-3\%}$, respectively (Table~\ref{fbarMV_tab}).}

\item{\textit{Optical bar fraction as a function of morphology and luminosity:} 

~We explore  $f_{\rm bar-opt}$ as a function of host galaxy properties and 
find that it rises in spiral galaxies, which  are less bulge-dominated and/or are
brighter. 
The optical bar fraction is a factor of $\sim$~1.8
higher in galaxies classified as `pure disk' compared to galaxies 
visually classified as `bulge+disk' (Table~\ref{fbarmorph_tab}). {\it Within a given  $M_{\rm V}$  bin,  $f_{\rm bar-opt}$ is higher in
visually-selected disk galaxies that have no bulge as opposed to those
with bulges}. Furthermore, we find that for a given visual morphological class,  $f_{\rm bar-opt}$ 
rises at higher absolute magnitudes (Fig.~\ref{fbarlum} and Table~\ref{fbarMVmorph_tab}).
When the normalized effective radius
 $r_{\rm e}/a_{\rm disk}$ is 
used to trace central galaxy concentration, the bar fraction is
$\sim$~2.7 times higher in galaxies with the lowest central   
concentration ($r_{\rm e}/a_{\rm disk} = 0.75$) compared to the   
galaxies with the highest central concentration ($r_{\rm e}/a_{\rm
  disk} = 0.15$; Fig.~\ref{bfBTre}).
 }

\item{\textit{Optical bar fraction as a function of $\kappa$, $\Sigma_{10}$,
    ICM density, and distance from nearest cluster center:} 
Between the core and the virial radius of the cluster ($R\sim$~0.25 to
    1.2 Mpc) at intermediate densities (log($\Sigma_{10}$)$ = 1.7$ to 2.3), 
the optical bar fraction does not appear to depend strongly on the local environment density 
tracers  ($\kappa$, $\Sigma_{10}$, and ICM density), and varies at most by a factor of 
$\sim$~1.3 (Fig.~\ref{fbdens}. Inside the cluster core, within the caveats of limited number statistics, 
$f_{\rm bar-opt}$ does not show evidence for a variation larger than a factor of 1.5 
as a function of the three environmental indicators. Overall, our
    results suggest that the optical bar fraction is not strongly
    dependent on environment at intermediate densities (e.g., log($\Sigma_{10}$)$ = 1.7$ to 2.3). 
}

\item{\textit{Comparison to field studies:} We compare in Table~\ref{fbarfield_tab} our results to
  those for field samples, specifically MJ07 (OSUBSGS) and A09 (SDSS), where bar
  identification and characterization was done through ellipse-fitting. We find that
  for bright early Hubble types, as well as faint late-type
  systems with no evident bulge, the optical bar fraction in the 
  Abell 901/2  clusters is comparable within a factor of 1.2 
   to that of  field galaxies at lower redshifts ($z<0.04$).}

\item{\textit{Bar strength distribution in cluster and field:} We find
that in both the cluster and field, the highest ellipticity bars lie
in disk-dominated galaxies.}

\end{enumerate}             
 S.J. and I.M. acknowledge support from the National Aeronautics and Space
Administration (NASA) LTSA grant NAG5-13063, NSF grant AST-0607748,
and $HST$ grants G0-10395 from STScI, which is operated by
AURA, Inc., for NASA, under NAS5-26555.
E.\ F.\ B.\  and K.\ J.\ acknowledge support from the Deutsche
Forschungsgemeinschaft through the Emmy Noether Programme.
A. B. acknowledges support from DLR grant 50 OR 0404. 
D. H. M. acknowledges support from NASA  LTSA grant NAG5-13102 issued
through the office of Space Science. E.vK. and M.B. were supported by
the Austrian Science Foundation FWF under grand P18416.
C. Y. P. is grateful for support provided through STScI and NRC-HIA
Fellowship. C. W. acknowledges support from  an STFC Advanced Fellowship.
Support for STAGES was provided by the NASA through $HST$ grant G0-10395 from STScI, which is operated by
AURA, Inc., for NASA, under NAS5-26555. The STAGES team would like to
thank Hans-Walter Rix for his support.
This research has made use of NASA's Astrophysics Data System Service.



{}                            


%
\clearpage
\setcounter{table}{0}
\begin{deluxetable}{lcccc}
\tabletypesize{\scriptsize}
\tablewidth{0pt}
\tablecaption{Galaxy Morphology as a Function of Distance from Cluster Centers}
\tablehead{
\colhead { } &
\colhead {Whole cluster sample}&
\colhead {Core}&
\colhead {Outer region}&
\colhead {Outskirt}\\ 
\colhead { } &
\colhead { } &
\colhead {R $<$ R$_{\rm core}$} & 
\colhead {R$_{\rm core} <$ R $<$ R$_{\rm virial}$} &
\colhead {R $>$ R$_{\rm virial}$}\\
\colhead { } &
\colhead {(1)}&
\colhead {(2)}&
\colhead {(3)}&
\colhead {(4)}\\
}
\startdata
N$_{\rm all}$ & 750 & 81  & 556  &  113\\
\hline \\
N$_{\rm bulge}$ & 125 & 26 & 88 & 11 \\
N$_{\rm disk}$  & 625 & 55  & 468  & 102  \\
N$_{\rm disk}$ (bulge+disk) & 452  & 46  & 334  & 72 \\
N$_{\rm disk}$ (pure disk)  & 173  & 9 & 134  & 30 \\
\hline \\
N$_{\rm bulge}$/N$_{\rm all}$& 0.17  &  0.32 & 0.16 & 0.10\\
N$_{\rm disk}$/N$_{\rm all}$&  0.83 &0.68  & 0.84  & 0.90\\
N$_{\rm bulge+disk}$/N$_{\rm all}$& 0.60  & 0.57  & 0.60 & 0.64 \\
N$_{\rm pure ~disk}$/N$_{\rm all}$& 0.23 & 0.11  & 0.24 & 0.27\\
\enddata
\tablecomments{The numbers shown are for the 750 bright ($M_{\rm V}
  \le -18$) galaxies, which we were able to classify into three broad
  groups: `pure bulge', `bulge+disk', and `pure disk', as outlined in
  $\S$~\ref{vcmeth}. The relative numbers in each group are shown for
  the whole cluster (column 1) and different regions within the custer
  (columns 2, 3, and 4). The core radius R$_{\rm core} = 0.25$~Mpc and
  the virial radius R$_{\rm virial} = 1.2$~Mpc. 
\label{morphdens_tab} }
\end{deluxetable}

\clearpage
\setcounter{table}{1}
\begin{deluxetable}{lccc}
\tabletypesize{\scriptsize}
\tablewidth{0pt}
\tablecaption{Optical Bar Fraction from Different Methods to Identify
  Disk Galaxies Among $M_{\rm V} \le -18$, $i<60^{\circ}$ Systems}
\tablehead{
\colhead {Method} &
\colhead {N$_{\rm disk}$} & 
\colhead {N$_{\rm barred}$} &
\colhead {f$_{\rm bar,opt}$}\\
\colhead {(1)}&
\colhead {(2)}&
\colhead {(3)}&
\colhead {(4)}\\
}
\startdata
Visual & 340 & 115 & 34\%$^{+10\%}_{-3\%}$ \\
Color  & 189 & 58  & 31\%$^{+10\%}_{-3\%}$  \\
S{\'e}rsic & 241 & 72 & 30\%$^{+10\%}_{-3\%}$ \\
\enddata
\tablecomments{All optical bar fractions are for galaxies with $M_{\rm V} \le -18$, $i<60^{\circ}$, and $a_{\rm disk}> 3$~kpc. 
Columns are: 
(1)~Method for selecting disk galaxies. See $\S$~\ref{diskselmeth} and $\S$~\ref{cldisks} for details;
(2)~Number of moderately inclined disk galaxies, N$_{\rm disk}$;
(3)~Number of barred disk galaxies, N$_{\rm barred}$. Bars are detected through
  ellipse fitting;
(4)~Optical bar fraction, $f_{\rm bar-opt}$, defined as in
  Eq. 1. The upper error bar on the optical bar fraction is the sum in quadrature of 
the error in the bar fraction from isophotal twists ($\S$~\ref{barchar}) and 
the statistical error. Note that including isophotal twists into the optical bar fraction can 
only make the optical bar fraction higher. Therefore, the lower error bars quoted represent only the statistical error.
\label{fbarMV_tab} }
\end{deluxetable}

\setcounter{table}{2}
\begin{deluxetable}{lccc}
\tabletypesize{\scriptsize}
\tablewidth{0pt}
\tablecaption{Optical Bar Fraction from Different Methods to Identify
  Disk Galaxies Among M$_{*}$/M$_{\odot} \ge 10^{9}$, $i<60^{\circ}$ Systems}
\tablehead{
\colhead {Method} &
\colhead {N$_{\rm disk}$} & 
\colhead {N$_{\rm bar}$} &
\colhead {f$_{\rm bar,opt}$}\\
\colhead {(1)}&
\colhead {(2)}&
\colhead {(3)}&
\colhead {(4)}\\
}
\startdata
Visual & 389 & 128 & 33\%$^{+10\%}_{-2\%}$  \\
Color  & 208 & 69  & 33\%$^{+10\%}_{-3\%}$  \\
S{\'e}rsic & 290 & 85 & 29\%$^{+10\%}_{-3\%}$ \\
\enddata
\tablecomments{All optical bar fractions are for galaxies with M$_{*}$/M$_{\odot}> 10^{9}$, 
$i<60^{\circ}$, and $a_{\rm disk} > 3$~kpc.  
Columns are: 
(1)~Method for selecting disk galaxies. See $\S$~\ref{diskselmeth} and $\S$~\ref{cldisks} for details;
(2)~Number of moderately inclined  disk galaxies, N$_{\rm disk}$;
(3)~Number of barred disk galaxies, N$_{\rm barred}$. Bars are detected through
  ellipse fitting;
(4)~Optical bar fraction, $f_{\rm bar-opt}$, defined as in Eq. 1. The upper error bar on the optical bar fraction is the sum in quadrature of the error in the bar fraction from isophotal twists ($\S$~\ref{barchar}) and 
the statistical error. Note that including isophotal twists into the optical bar fraction can 
only make the optical bar fraction higher. Therefore, the lower error bars quoted represent only 
the statistical error.
\label{fbarM_tab}}
\end{deluxetable}

\setcounter{table}{3}
\begin{deluxetable}{lcccc}
\tabletypesize{\scriptsize}
\tablewidth{0pt}
\tablecaption{Optical Bar Fraction as a Function of Visually
  Classified Secondary Morphological Parameters}
\tablehead{
\colhead {Morphology } &
\colhead {N$_{\rm all}$} & 
\colhead {N$_{\rm disk}$} &
\colhead {N$_{\rm bar}$}&
\colhead {f$_{\rm bar,opt}$}\\
\colhead {(1)}&
\colhead {(2)}&
\colhead {(3)}&
\colhead {(4)}&
\colhead {(5)}\\
}
\startdata
Pure bulge & 105 & --- & --- & ---  \\
B+D    & 110$^{a}$+131$^{b}$+21$^{c}$ & 262  & 77 & 29\%$^{+10\%}_{-3\%}$ \\
Pure disk  & 78 & 78 & 38 & 49\%$^{+12\%}_{-6\%}$ \\
\hline \\
Clumpy     & 105 & 105 & 47 & 45\%$^{+11\%}_{-5\%}$ \\
Smooth     & 340 & 235 & 68 & 29\%$^{+10\%}_{-3\%}$ \\
\enddata
\tablecomments{All values are for galaxies with $M_{\rm V} \le -18$, $i<60^{\circ}$, and $a_{\rm max} > 3$~kpc.
 Columns are~: 
(1)~Morphological parameters from visual classification ($a$ - number of
  `bulge+disk' galaxies with bar/spiral arm; $b$ - number of `bulge+disk'
  galaxies without bar/spiral; $c$ - number of bulge+disk galaxies
  without bar and no spiral arm class);
(2)~Total number of galaxies in class; 
(3)~Number of moderately inclined \textit{disk} galaxies in class; 
(4)~Number of barred disk galaxies, where bars are from ellipse fitting; 
(5)~Optical bar fraction calculated as in Eq. 1. The upper error bar on the optical bar 
fraction is the sum in quadrature of the error in the bar fraction from isophotal twists 
($\S$~\ref{barchar}) and the statistical error. Note that including isophotal twists 
into the optical bar fraction can only make the optical bar fraction higher. Therefore, 
the lower error bars quoted represent only the statistical error.
\label{fbarmorph_tab}}
\end{deluxetable}

\setcounter{table}{4}
\begin{deluxetable}{lcccc}
\tabletypesize{\scriptsize}
\tablewidth{0pt}
\tablecaption{Optical Bar Fraction as a Function of Host Absolute
Magnitude and Morphological Class}
\tablehead{
\colhead {$M_{\rm V}$ range} &
\colhead {Bulge+Disk Smooth} & 
\colhead {Bulge+Disk Clumpy} &
\colhead {Pure Disk Smooth}&
\colhead {Pure Disk Clumpy}\\
}
\startdata
$-18 \ge M_{\rm V} > -19$& 10\%$\pm4\%$ (6/58) & 40\%$\pm22\%$ (2/5) & 46\%$\pm10\%$ (11/24) & 29\%$\pm11\%$ (5/17)  \\
$-19 \ge M_{\rm V} > -20$& 20\%$\pm5\%$ (14/71) & 18\%$\pm9\%$ (3/17) & 57\%$\pm19\%$ (4/7) & 60\%$\pm11\%$ (12/20)  \\
$-20 \ge M_{\rm V} > -21$& 42\%$\pm7\%$ (22/53) & 35\%$\pm11\%$ (7/20) & --- & 63\%$\pm17\%$ (5/8)  \\
$-21 \ge M_{\rm V} > -22$& 53\%$\pm11\%$ (10/19) & 75\%$\pm11\%$ (12/16) & --- & ---  \\
\enddata
\tablecomments{We show the variation of $f_{\rm bar-opt}$ as a function of 
absolute magnitude and morphological class for visually-identified, moderately inclined
disk galaxies with $a_{\rm disk} > 3$~kpc.  The numbers in parentheses give the values
$N_{barred}/(N_{barred}+N_{unbarred})$ in each bin. Values are only shown for bins containing more than 
two galaxies. The error presented is the statistical error in each bin. 
\label{fbarMVmorph_tab} }
\end{deluxetable}

\setcounter{table}{5}
\begin{deluxetable}{lcccc}
\tabletypesize{\scriptsize}
\tablewidth{0pt}
\tablecaption{Optical Bar Fraction as a Function of $U-V$ Color and Absolute Magnitude}
\tablehead{
\colhead { } &
\colhead {$-18 \ge M_{\rm V} > -19$} & 
\colhead {$-19 \ge M_{\rm V} > -20$} &
\colhead {$-20 \ge M_{\rm V} > -21$}&
\colhead {$-21 \ge M_{\rm V} > -22$}\\
}
\startdata
$U-V < 1$& 31\%$\pm6\%$ (20/65) & 34\%$\pm7\%$ (17/50) & 36\%$\pm10\%$ (8/22) & 20\%$\pm18\%$ (1/5)  \\
$U-V > 1$& 10\%$\pm5\%$ (4/39) & 25\%$\pm5\%$ (16/65) & 44\%$\pm6\%$ (26/59) & 69\%$\pm8\%$ (22/32)  \\
\enddata
\tablecomments{We show the variation of $f_{\rm bar-opt}$ as a function of 
rest-frame $U-V$ color  and absolute magnitude for visually-identified, moderately inclined
disk galaxies with $a_{\rm disk} > 3$~kpc.  The numbers in parentheses give the values
$N_{barred}/(N_{barred}+N_{unbarred})$ in each bin. Values are only shown for bins containing more than 
two galaxies. The error presented is the statistical error in each bin. 
\label{fbarMVcolor_tab}}
\end{deluxetable}

\setcounter{table}{6}
\begin{deluxetable}{lcccc}
\tabletypesize{\scriptsize}
\tablewidth{0pt}
\tablecaption{Optical Bar Fraction as a Function of $U-V$ Color and Morphological Class}
\tablehead{
\colhead { } &
\colhead {Bulge+Disk Smooth} & 
\colhead {Bulge+Disk Clumpy} &
\colhead {Pure Disk Smooth}&
\colhead {Pure Disk Clumpy}\\
}
\startdata
$U-V < 1$& 6\%$\pm4\%$ (2/33) & 21\%$\pm7\%$ (8/38) & 52\%$\pm10\%$ (14/27) & 49\%$\pm7\%$ (22/45)  \\
$U-V > 1$& 30\%$\pm4\%$ (51/170) & 76\%$\pm9\%$ (16/21) & 20\%$\pm18\%$ (1/5) & ---  \\
\enddata
\tablecomments{We show the variation of $f_{\rm bar-opt}$ as a function of 
rest-frame $U-V$ color  and morphological class for visually-identified, moderately inclined
disk galaxies with $a_{\rm disk} > 3$~kpc.  The numbers in parentheses give the values
$N_{barred}/(N_{barred}+N_{unbarred})$ in each bin. Values are only shown for bins containing more than 
two galaxies. The error presented is the statistical error in each bin. 
\label{fbarcolormorph_tab} }
\end{deluxetable}

\setcounter{table}{7}
\begin{deluxetable}{lcc}
\tabletypesize{\scriptsize}
\tablewidth{0pt}
\tablecaption{Comparison of local optical bar fraction in field and clusters}
\tablehead{
\colhead { } &
\colhead { } & 
\colhead { } \\
}
\startdata
\multicolumn {3}{c} {\small  A) Bright galaxies split into morphological types} \\
\hline \\
Reference & MJ07 & this work  \\
Magnitude range & $M_{\rm V}\le -20$ & $M_{\rm V}\le -20$ \\
Environment    & OSUBSGS (mostly field) & Abell 901/902 (cluster) \\
Redshift  & $\sim$~0 & 0.165  \\
$f_{\rm bar,opt}^{1}$ & All types$^{2}$: 40\%$^{+8\%}_{-5\%}$ (36/90) & All types: 48\%$^{+11\%}_{-5\%}$ (58/121)\\
                  & S0-Sbc$^{2}$: 41\%$^{+10\%}_{-6\%}$ (31/71) & B+D: 47\%$^{+11\%}_{-5\%}$ (52/111)\\
\hline \\
\hline \\
 & & \\
\multicolumn {3}{c} {\small B) Bright galaxies split into morphological types} \\
\hline \\
Reference & A09 & this work  \\
Magnitude range & $M_{\rm V}~\le~-20.5$ & $M_{\rm V}~\le~-20.5$ \\
Environment    & SDSS (field + interm. density) & Abell 901/902 (cluster) \\
Redshift  & 0.01-0.04 & 0.165  \\
$f_{\rm bar,opt}$ & All types: 48\% (337/699) & All types$^{3}$: 57\%$^{+12\%}_{-6\%}$ (41/72)\\
                  & S0 to Sbc: 41\% (196/474)  & B+D $^{3}$: 56\%$^{+12\%}_{-6\%}$ (39/68)\\
\hline \\
\hline \\
 & & \\
\multicolumn {3}{c} {\small C) Faint galaxies split into morphological types} \\
\hline \\
Reference & A09 & this work  \\
Magnitude range & $-20.5~\le~M_{\rm V}~\le~-18.6$ & $-20.5~\le~M_{\rm V}~\le~-18.6$ \\
Environment    & SDSS (field + interm. density) & Abell 901/902 (cluster)  \\
Redshift  & 0.01-0.04 & 0.165  \\
$f_{\rm bar,opt}$ & All types: 43\% (589/1360) & All types: 32\%$^{+10\%}_{-3\%}$ (65/205)\\
                  &  S0 to Sbc: 39\% (349/893) & B+D: 23\%$^{+10\%}_{-4\%}$ (37/159)\\
                  &  Scd-Sd: 51\% (240/467) & pure disk: 61\%$^{+12\%}_{-7\%}$ (28/46) \\
\hline \\
\hline \\
\enddata
\tablecomments{(1): The fractions quoted for this study are for galaxies with $i<60^{\circ}$ and $a_{\rm disk} > 3$~kpc.
(2): The upper error bar on the optical bar fraction quoted for this study and  MJ07
is the sum in quadrature of the error in the bar fraction from isophotal twists ($\S$~\ref{barchar}) and 
the statistical error. Note that including isophotal twists into the optical bar fraction can 
only make the optical bar fraction higher. Therefore, the lower error bars quoted represent only the statistical error. \\
(3):~The OSUBSGS sample of moderately inclined galaxies in MJ07 
is  dominated by early-to-intermediate Hubble types (S0-Sbc; 71) galaxies 
mostly S0/a to Sbc. The number of late Hubble type (Sc-Sm; 18) galaxies is 
too low to yield robust number statistics for the late types. We thus only
show $f_{\rm bar,opt}$ for the early-to-intermediate Hubble types.\\
(4):~In the STAGES sample, there are only 5 pure disk galaxies with magnitudes
$M_{\rm V} \le -20.5$, while most pure disk galaxies have $M_{\rm V} \ge -20$ (see Table~\ref{fbarMVmorph_tab}).  
For this reason, we only show the bright-galaxy comparison for early-type (B+D) galaxies. 
\label{fbarfield_tab}}
\end{deluxetable}

\clearpage
\setcounter{figure}{0}
\begin{figure}
\epsscale{0.7}
\plotone{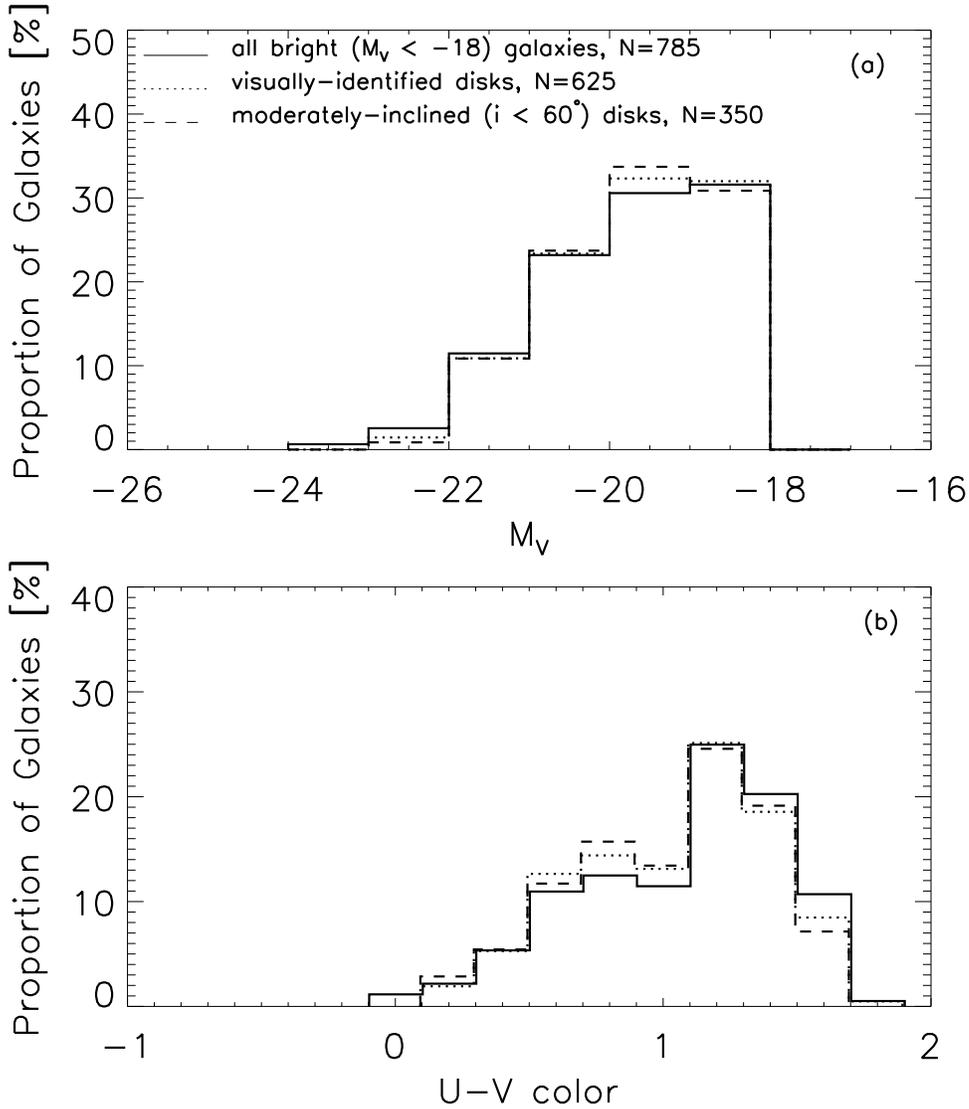}
\vskip 1 in 
\caption{
(a)~The solid line shows the histogram of  absolute magnitude $M_{\rm V}$ of our total 
cluster sample of 785 bright ($M~_{\rm V} \le -18$)
galaxies. Most galaxies have $-20 \le M_{\rm V} \le -18$. The dotted
line shows the M$_{\rm V}$ distribution of galaxies visually classified as disks. 
The dashed line shows the M$_{\rm V}$ distribution of the final ellipse-fitted disk sample,
after excluding highly inclined ($i > 60^{\circ}$), and poorly fitted
galaxies. (b)~Rest-frame $U-V$ color
distribution of the whole cluster galaxy sample (solid line), visually-identified 
disk sample (dotted line), and ellipse-fitted, moderately-inclined disk sample (dashed line).  
 Excluding highly inclined disk galaxies does not have a significant effect on the absolute $M_{\rm
  V}$ magnitude, or rest-frame $U-V$ color distributions. 
\label{MVcolordist}} 
\end{figure}

\clearpage
\begin{figure}
\vskip 0.2 in
\setcounter{figure}{1}
\epsscale{0.6}
\plotone{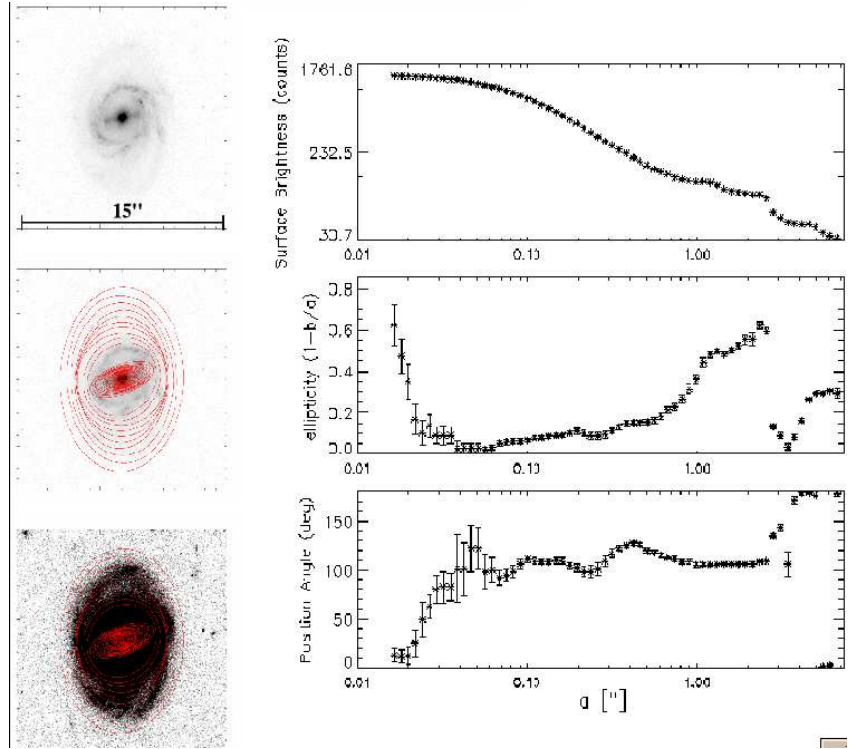}
\caption{
\textit{Left}: Ellipse fit overlays on the F606W image of a barred cluster
galaxy. In the middle and bottom panels, the contrast is adjusted to
show the inner regions and outer disk regions, respectively. \textit{Right}:
Radial profiles of the surface brightness (SB), ellipticity $e$, and
position angle (PA). The bar signature is evident
in the smooth rise of the $e$ to a global maximum, while the PA
remains relatively constant in the bar region. The $e$ then drops and
the PA changes, indicating the transition to the disk region. See
$\S$~\ref{barchar} for details.
\label{overlay}}
\end{figure}

\clearpage
\begin{figure}
\vskip 0.2 in
\setcounter{figure}{2}
\epsscale{0.7}
\plotone{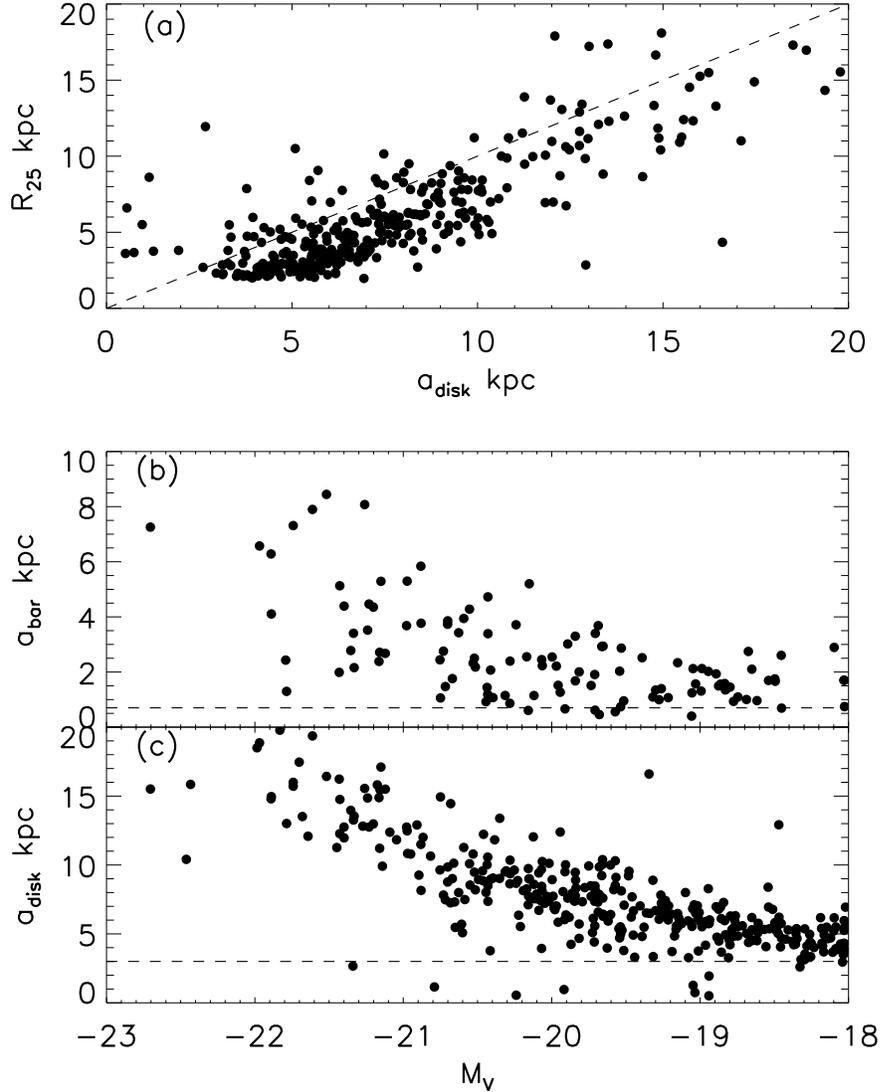}
\caption{
\textit{(a)}: The semi-major axes $a_{\rm disk}$ of galaxies visually classified
as disks ($\S$~\ref{diskselmeth}) correlate with the isophotal radius $R_{25}$ 
where the $B$-band surface brightness reaches 25~mag~arcsec$^{-2}$. The mean
ratio of $R_{25}$/$a_{\rm disk} = $~0.87. The dashed line shows 
a slope of 1. 
\textit{(b)}: The 
relationship between bar semi-major axis length $a_{\rm bar}$ and $M_{\rm V}$ absolute 
magnitude. The dashed line shows the limit of $a_{\rm bar} \sim~700$~pc
for reliable bar detection and characterization  using ellipse-fit.
\textit{(c)}: The 
relationship between disk semi-major axis length $a_{\rm disk}$ and absolute 
magnitude $M_{\rm V}$. For the bright $M_{\rm V} \le -18$ sample, we
only select disks with $a_{\rm disk} \ge 3$~kpc in order to ensure that
the bars of interest typically have $a_{\rm bar} \ge 700$~pc and can
be reliably detected. See $\S$~\ref{barchar} for details. 
\label{sizes}}
\end{figure}

\clearpage
\begin{figure}
\vskip 0.2 in
\setcounter{figure}{3}
\epsscale{1}
\plotone{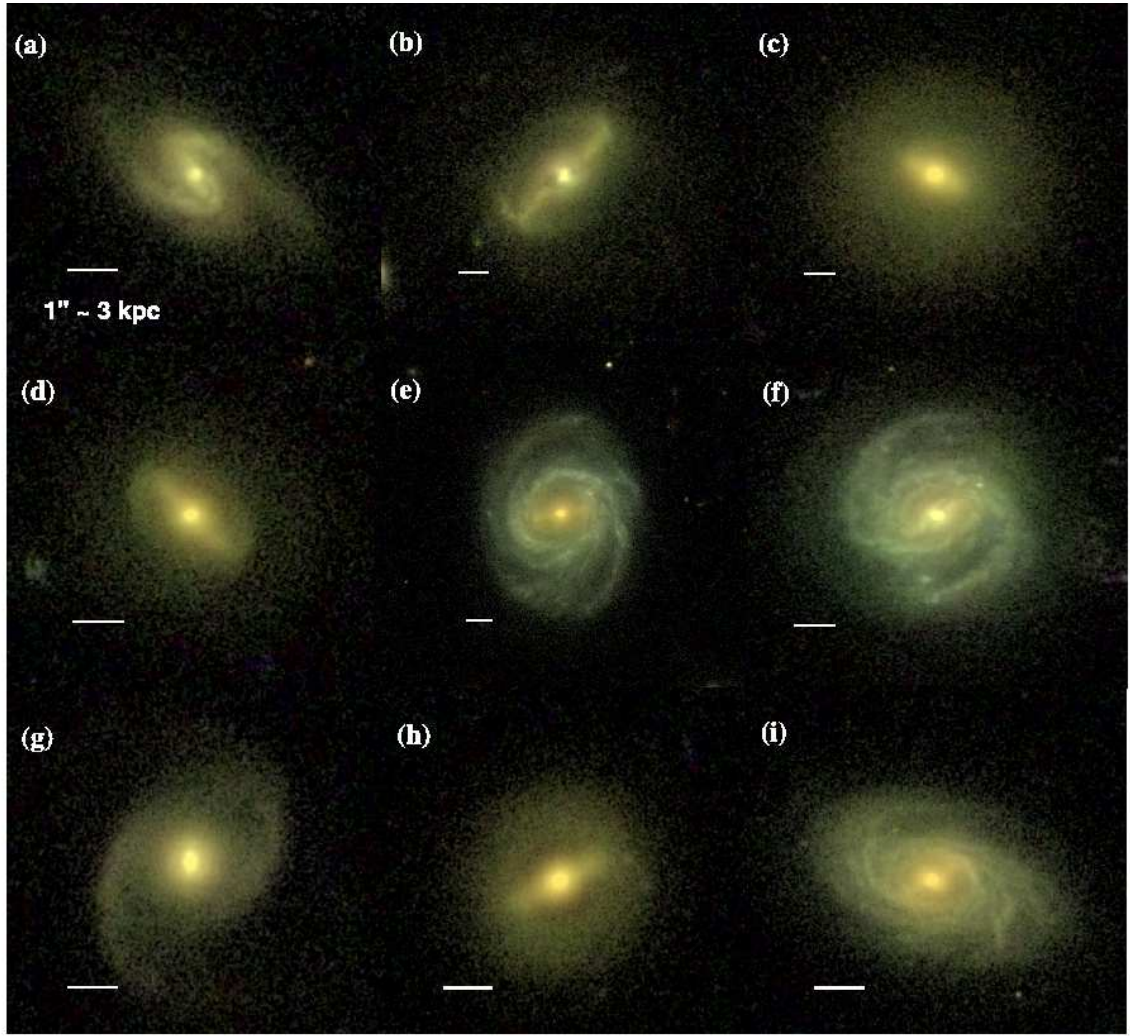}
\caption{Examples of representative bright ($M_{\rm V} \le -18$) barred galaxies identified through
  ellipse-fitting in the A901/902 supercluster.  The white 
line in each panel shows the scale of 1$\arcsec$~$\sim$~3~kpc.
\label{bars}}
\end{figure}

\clearpage
\begin{figure}
\vskip 0.2 in
\setcounter{figure}{4}
\epsscale{1}
\plotone{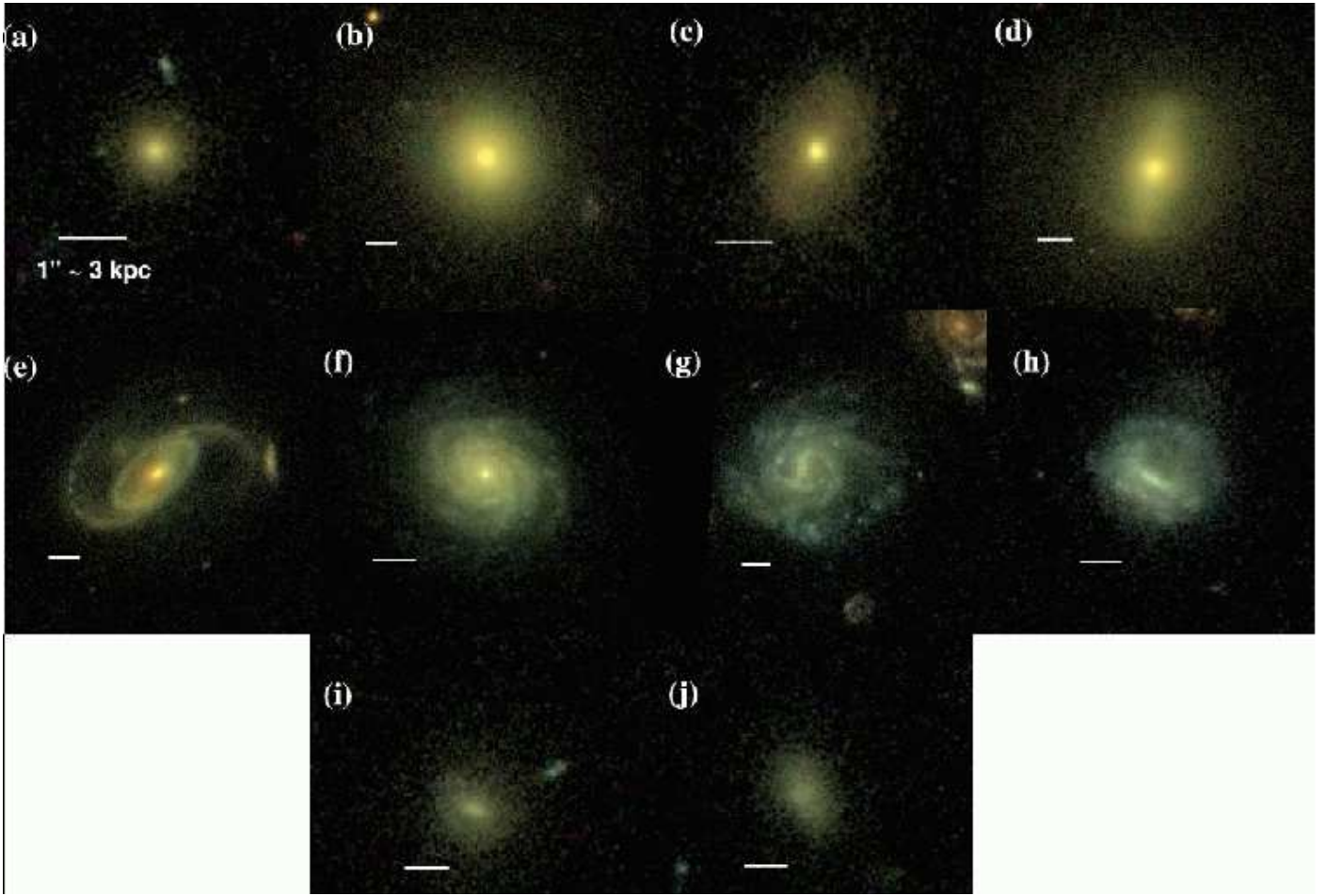}
\caption{Examples of the visual classification of secondary
  morphological properties ($\S$~\ref{vcmeth}) for the bright ($M_{\rm
  V} \le -18$), moderately inclined ($i< 60^{\circ}$) sample.  The white 
line in each panel shows the scale of 1$\arcsec$~$\sim$~3~kpc. Galaxies are grouped according
  to the visual prominence of the bulge into three groups: `pure
  bulge' (a,b), `bulge+disk' (c--f), and `pure disk' (g--j). Note that
  it is difficult to visually separate the classes `pure bulge' and
  `bulge+disk' (e.g., b vs. c) when the galaxy appears smooth and shows no disk
  signatures such as bars or spiral arms.
\label{morphex}}
\end{figure}

\clearpage
\begin{figure}
\setcounter{figure}{5}
\epsscale{0.7}
\plotone{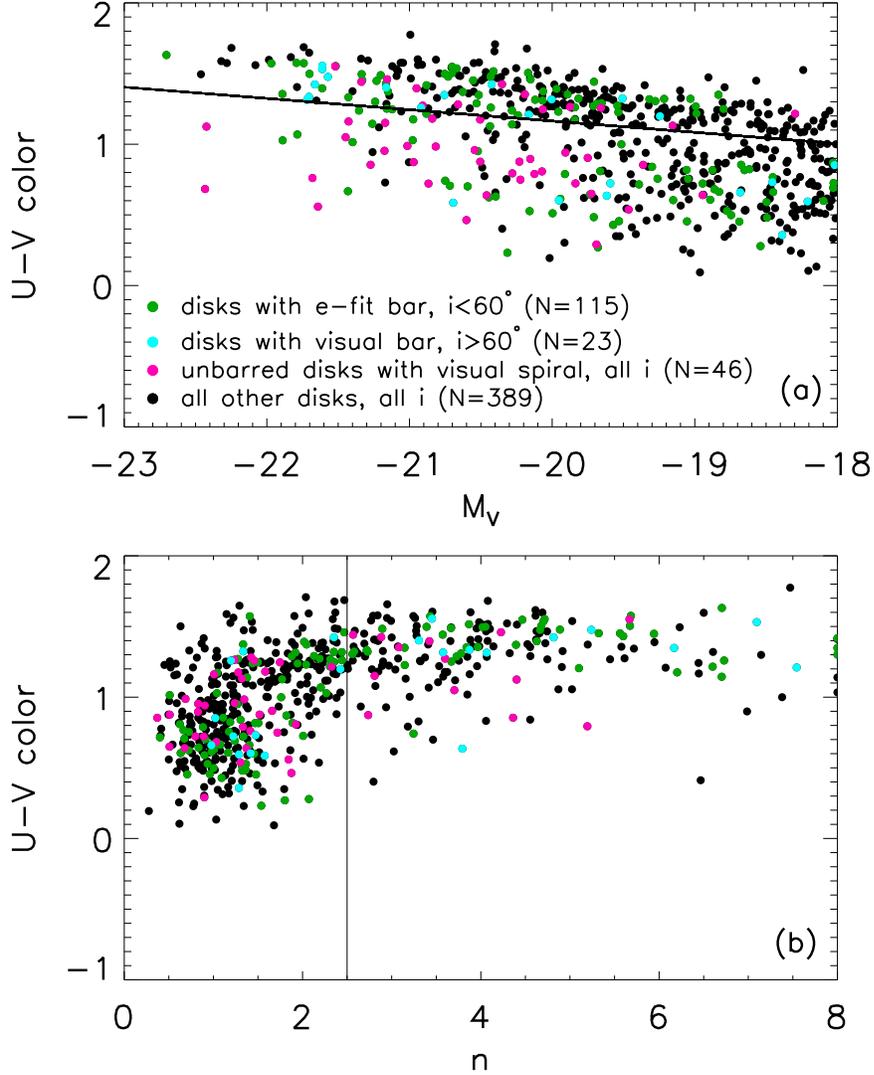}
\vskip 1 in 
\caption{
This figure compares the disk galaxies identified through three
different methods: visual classification, blue-cloud color cut, and a
S{\'e}rsic cut for the bright ($M_{\rm
  V} \le -18$) sample with $a_{\rm disk} > 3$~kpc. 
Panel (a) shows where the visually-identified disk galaxies 
lie in the rest-frame $U-V$ vs. $M_{\rm V}$ plane. Moderately-inclined,
$i < 60^{\circ}$, barred galaxies are
shown as green points, where the bars are identified through
ellipse-fitting. Bars in highly inclined galaxies  ($i > 60^{\circ}$), 
identified during visual classification are shown as cyan points. 
Unbarred disk galaxies with visually-identified spiral arms (all
inclinations) are shown in pink. The black points show galaxies
identified as disks with visual classification for all inclinations, but without a bar or
spiral arms.   The solid
line separates the red sample from the blue cloud galaxies. A blue-cloud color
cut selecting disks only below this line captures 279 out of 573
visually-identified disk galaxies. The remaining 294 or
$\sim$~51\%$\pm$2\% of visually-identified disks lie in the red sample. 
Panel (b) shows where visually identified disk galaxies lie in 
the S{\'e}rsic index $n$ vs. $M_{\rm V}$ plane. Colors are the same as in panel~(a).  
The solid line shows the cutoff of $n=2.5$, which is supposed to separate disk galaxies and
spheroids. Again, if such a cut is used to select disks (e.g., S0-Sm), 396 of the
visually-identified disks are captured, but the remaining 176 (31\%$\pm$2\%) with $n> 2.5$ are missed. 
\label{colsers}}
\end{figure}

\clearpage
\begin{figure}
\setcounter{figure}{6}
\epsscale{1}
\plotone{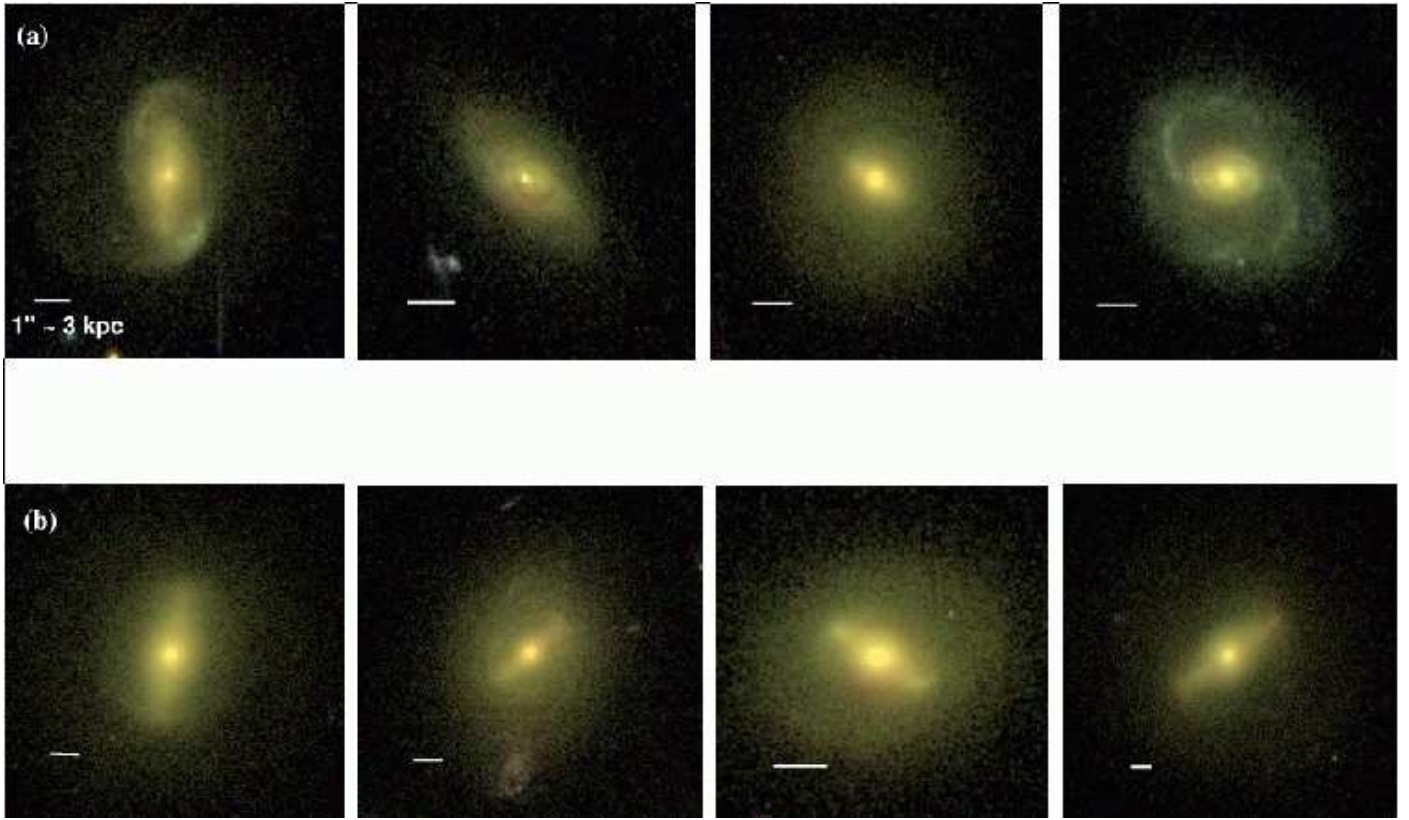}
\vskip 1 in 
\caption{
Examples of bright ($M_{\rm
  V} \le -18$), moderately inclined ($i< 60^{\circ}$), visually-identified disk galaxies,
which are missed by a S{\'e}rsic cut
with $n\le 2.5$ (a),  or by a blue-cloud cut (b).  The white 
line in each panel shows the scale of 1$\arcsec$~$\sim$~3~kpc.
\label{miss}}
\end{figure}

\clearpage
\begin{figure}
\setcounter{figure}{7}
\epsscale{0.7}
\plotone{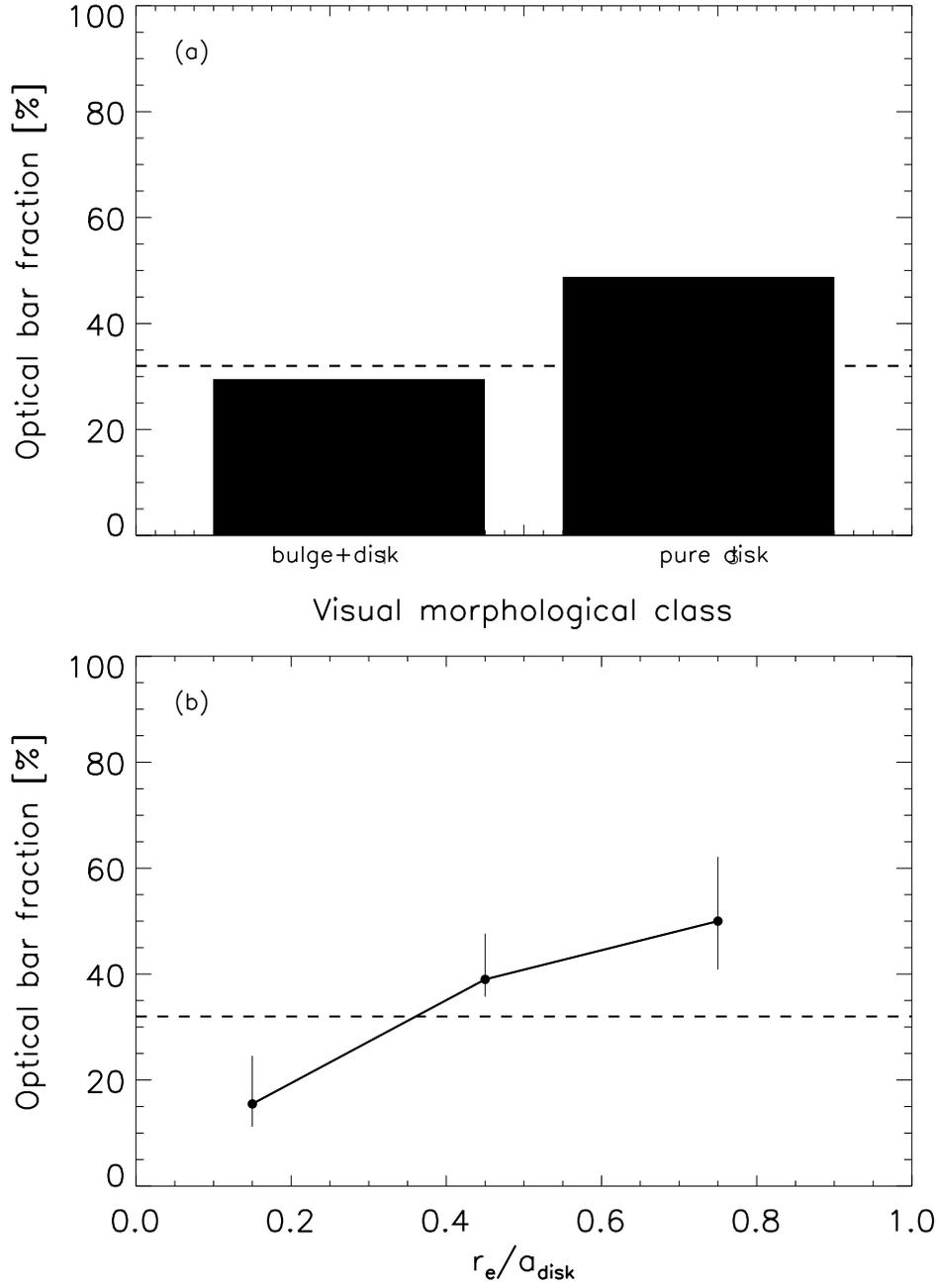}
\vskip 1 in 
\caption{\textit{(a)}~The optical bar fraction as a function of visual morphological
  class. The total bar fraction (34\%$^{+10\%}_{-3\%}$) based on disk galaxies of 
all morphological types using visual disk selection is
  shown as the horizontal dashed line in both panels. The first bin contains galaxies classified as `bulge+disk',
  while the second bin contains galaxies classified as `pure disk'. 
   The bar fraction shows a rise
  from 29\%$^{+10\%}_{-3\%}$ to 49\%$^{+12\%}_{-6\%}$ from galaxies classified as
  `bulge+disk' to `pure disk'.  \textit{(b)}~The optical bar fraction as a function of central galaxy
  concentration, as characterized by the effective radius normalized
  to the disk radius, $r_{\rm e}$/$a_{\rm disk}$. Only bins with
  significant number statistics are shown. The bar fraction
  increases from 15\%$^{+11\%}_{-4\%}$ in galaxies with high concentration
  ($r_{\rm e}/a_{\rm disk} \sim$~0.15), to 50\%$^{+14\%}_{-9\%}$ in galaxies with low
  concentration ($r_{\rm e}/a_{\rm disk}\sim$~0.75).  
\label{bfBTre}}
\end{figure}

\clearpage
\begin{figure}
\setcounter{figure}{8}
\epsscale{0.7}
\plotone{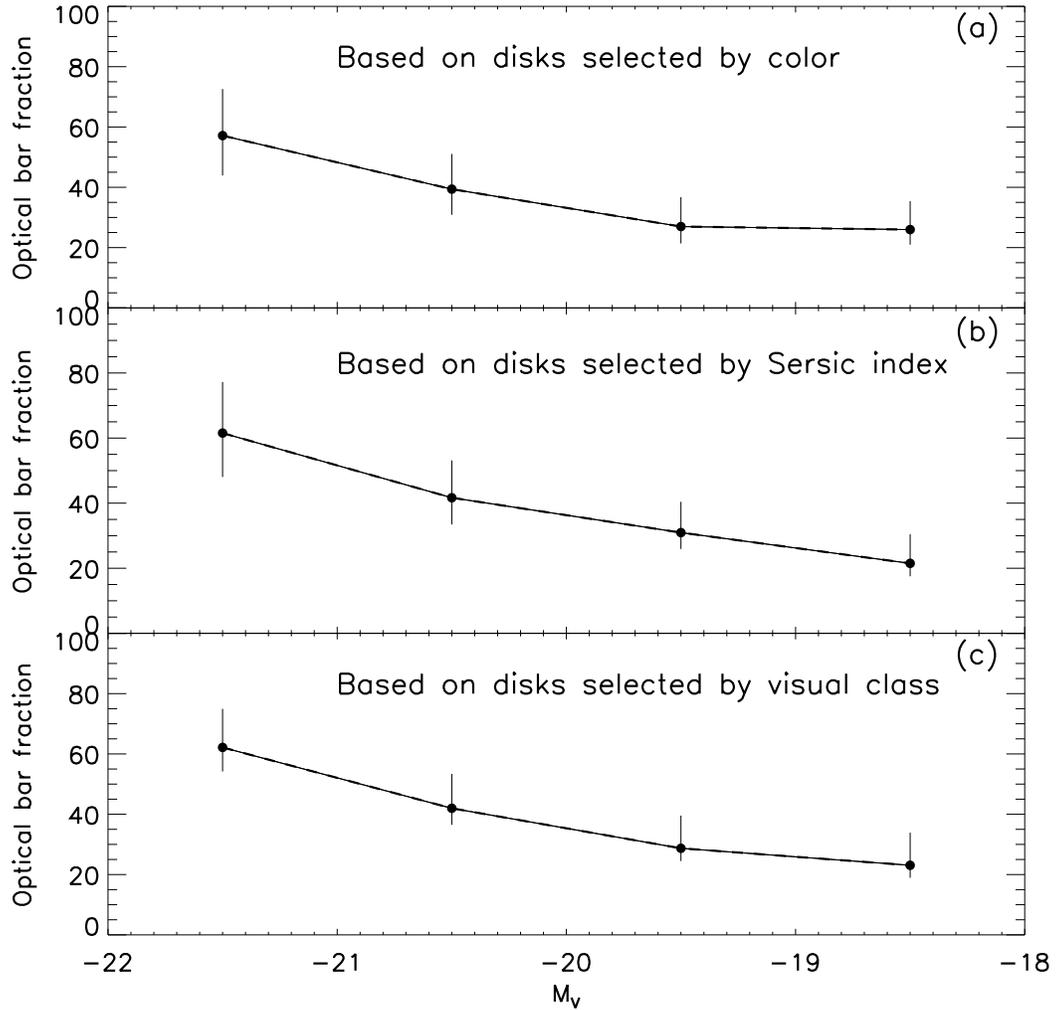}
\vskip 1 in 
\caption{~We plot the optical bar fraction as a function of galaxy 
luminosity $M_{\rm   V}$ for the three methods of disk selection:  
\textit{(a)}~a blue-cloud color cut; \textit{(b)}~a S{\'e}rsic ($n \le 2.5$) cut;  
\textit{(c)}~visual classification. 
For all three methods of disk selection, the optical bar fraction shows a decrease
from $\sim$~60\%$^{+14\%}_{-10\%}$ at $M_{\rm V} \sim -21.5$ to $\sim$~20\%$^{+11\%}_{-4\%}$ at 
$M_{\rm  V} = -18.5$. 
\label{fbarlum}}
\end{figure}

\clearpage
\begin{figure}
\setcounter{figure}{9}
\epsscale{0.7}
\plotone{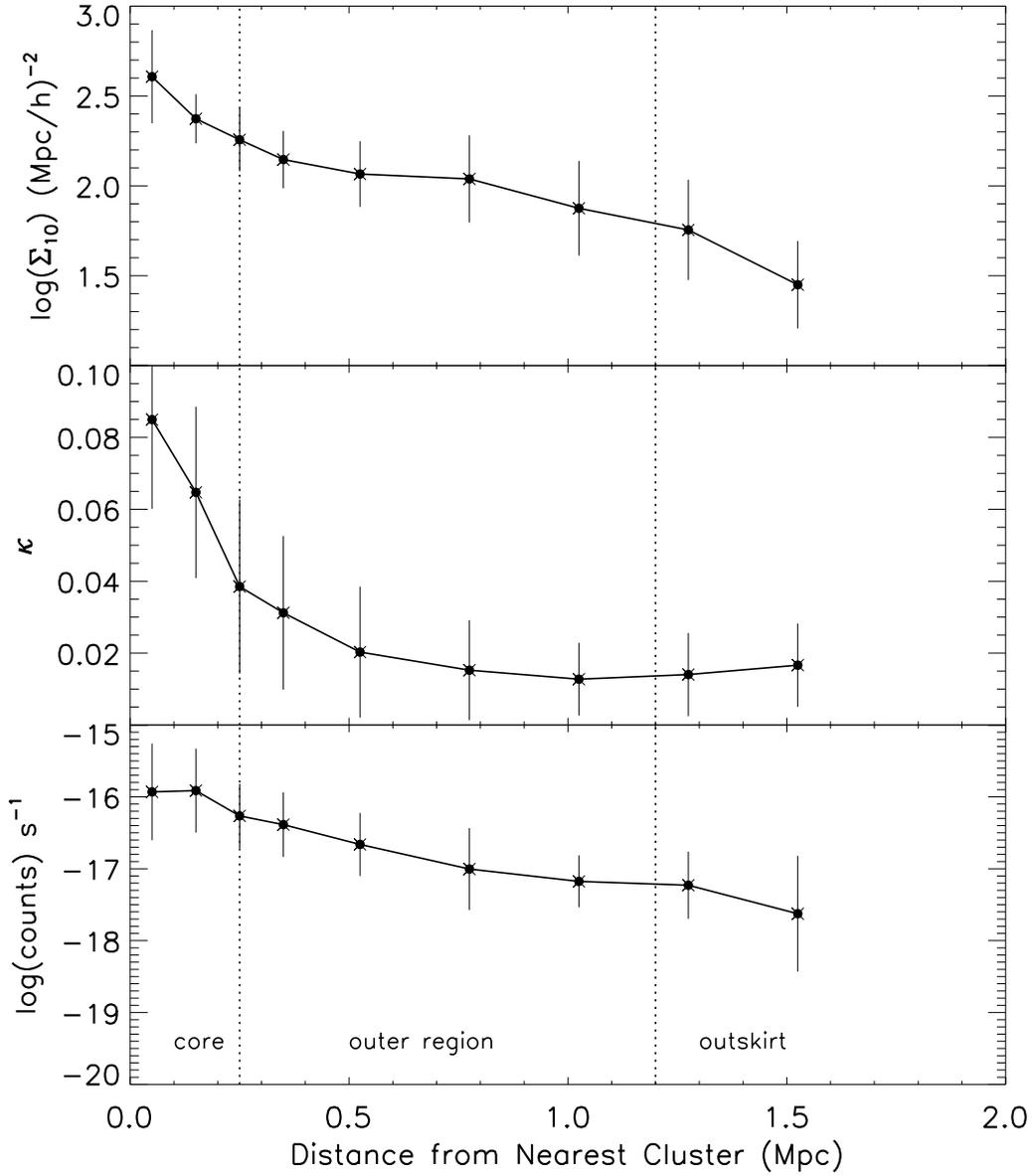}
\vskip 1 in
\caption{We plot the variation of the three measures of environment
  density ($\kappa$,  $\Sigma_{10}$, ICM density) as a function of 
  distance to the nearest cluster center. All three measures show a
  decrease in density as a function of cluster-centric distance. The
  vertical dashed lines denote the core radius at 0.25~Mpc and the
  virial radius at 1.2~Mpc. The error bars show the statistical
  Poisson errors in each bin. 
\label{stuffvdmin}}
\end{figure}

\clearpage
\begin{figure}
\setcounter{figure}{10}
\epsscale{1}
\plotone{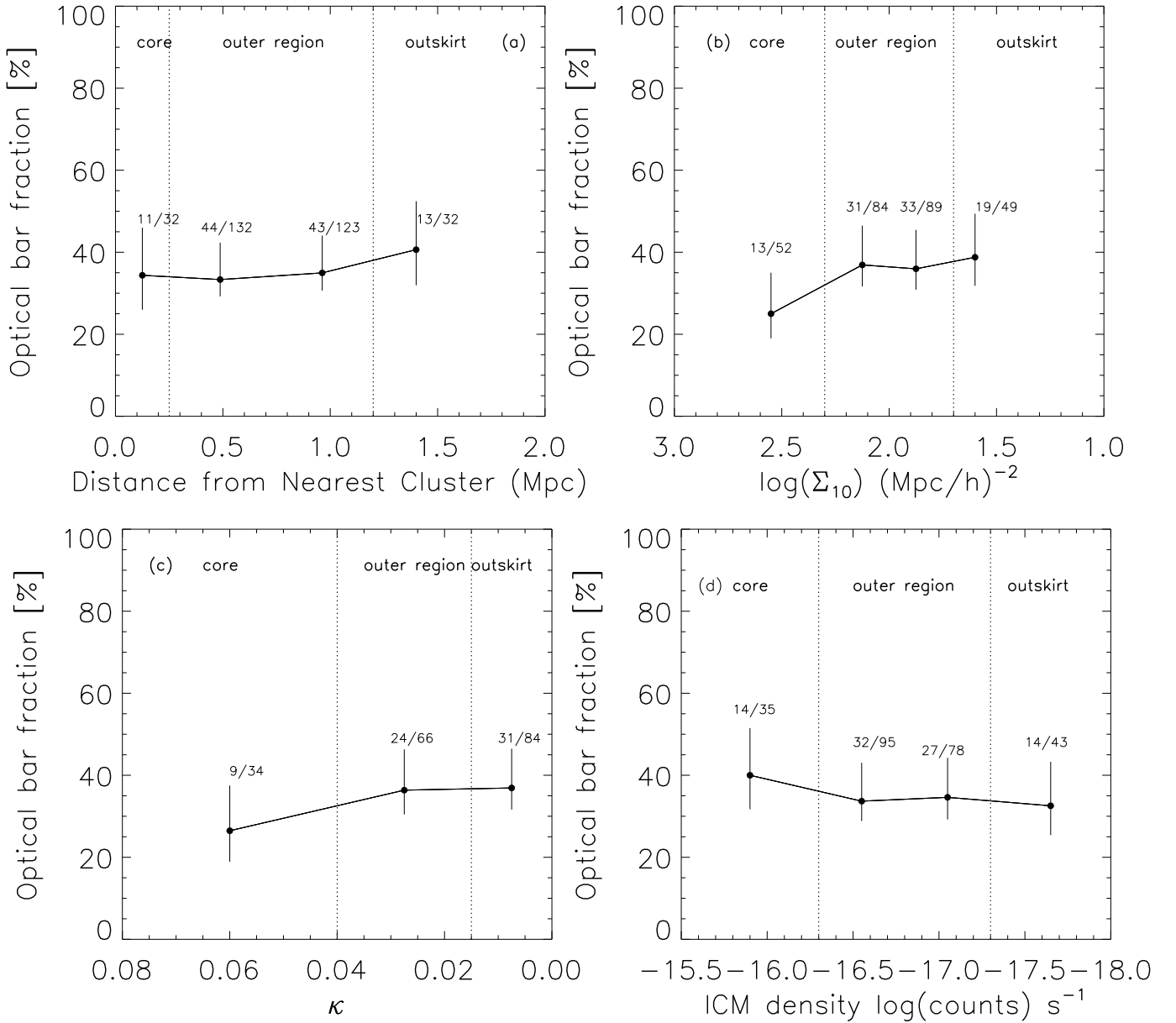}
\vskip -1 in 
\caption{The fraction of barred galaxies a function of:~ (a)~distance from nearest cluster
  center, (b)~log$\Sigma_{10}$, (c)~$\kappa$, and (d)~ICM
  density.  Bar classifications are from ellipse
  fits and disks  are identified by visual classification. The
  vertical dashed lines   denote the core radius at 0.25~Mpc and the
  virial radius at 1.2~Mpc. We find that between the core and the virial radius of the cluster 
($R\sim$~0.25 to 1.2 Mpc), the optical bar fraction $f_{\rm bar-opt}$ 
does not depend strongly on the local environment density tracers  ($\kappa$, 
$\Sigma_{10}$, and ICM density), and varies at most by a factor of $\sim$~1.3, 
allowed by the error bars.
\label{fbdens}}
\end{figure}

\clearpage
\begin{figure}
\setcounter{figure}{11}
\plotone{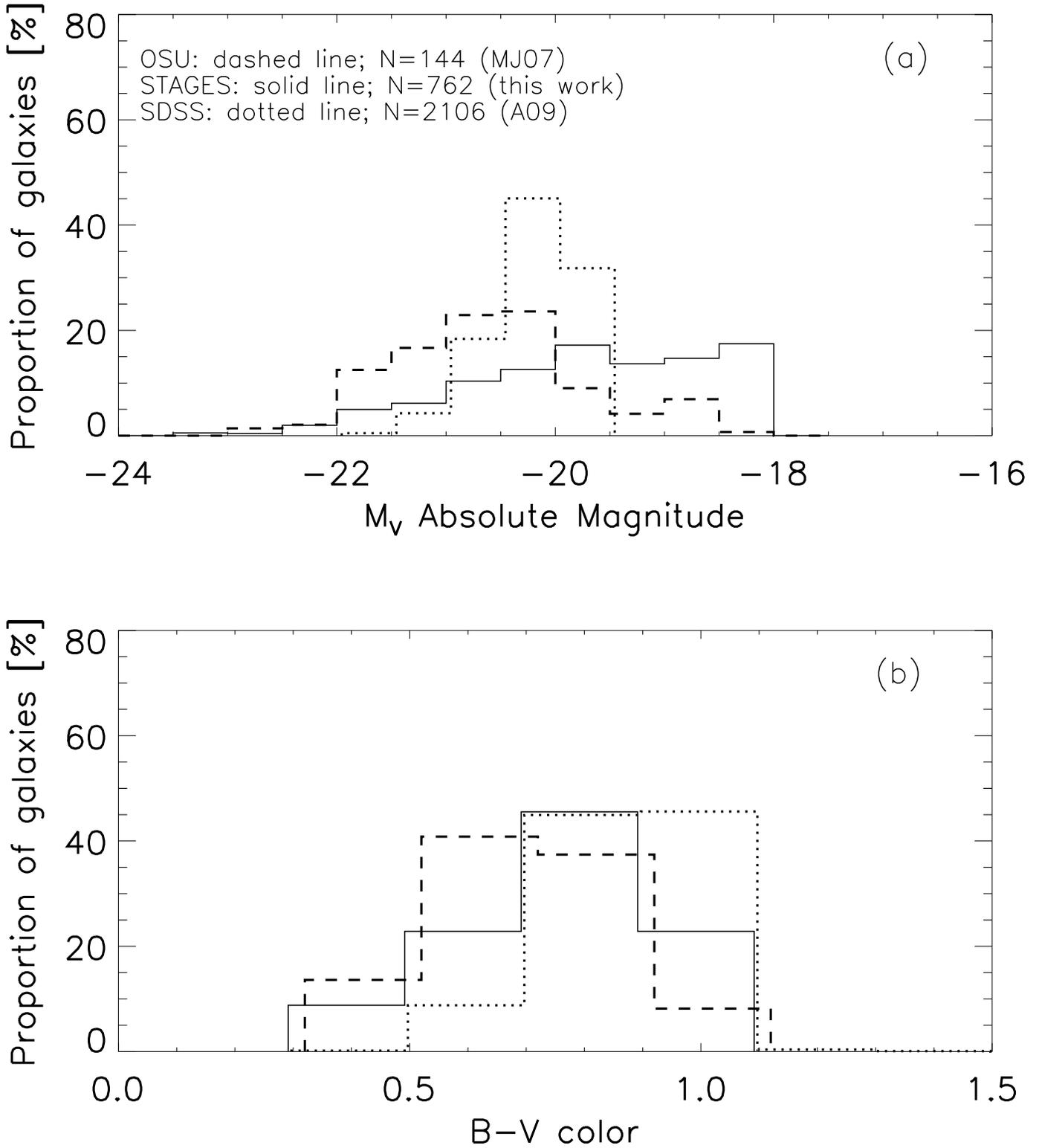}
\epsscale{0.7}
\vskip 1 in 
\caption{The absolute magnitude $M_{\rm V}$ \textit{(a)} and rest-frame $U-V$
  color \textit{(b)} distributions are shown for the OSUBSGS (dashed line), STAGES (solid line),
  and SDSS (dotted line) samples. The SDSS data are from A09. The
  OUSBSGS data are from MJ07. The OSUBSGS
  sample is brighter and  somewhat bluer
  than the STAGES sample. The SDSS sample spans a much narrower range in $M_{\rm V}$, with no galaxies
fainter than $-19.5$. 
\label{sampcolMV}}
\end{figure}

\clearpage
\begin{figure}
\setcounter{figure}{12}
\plotone{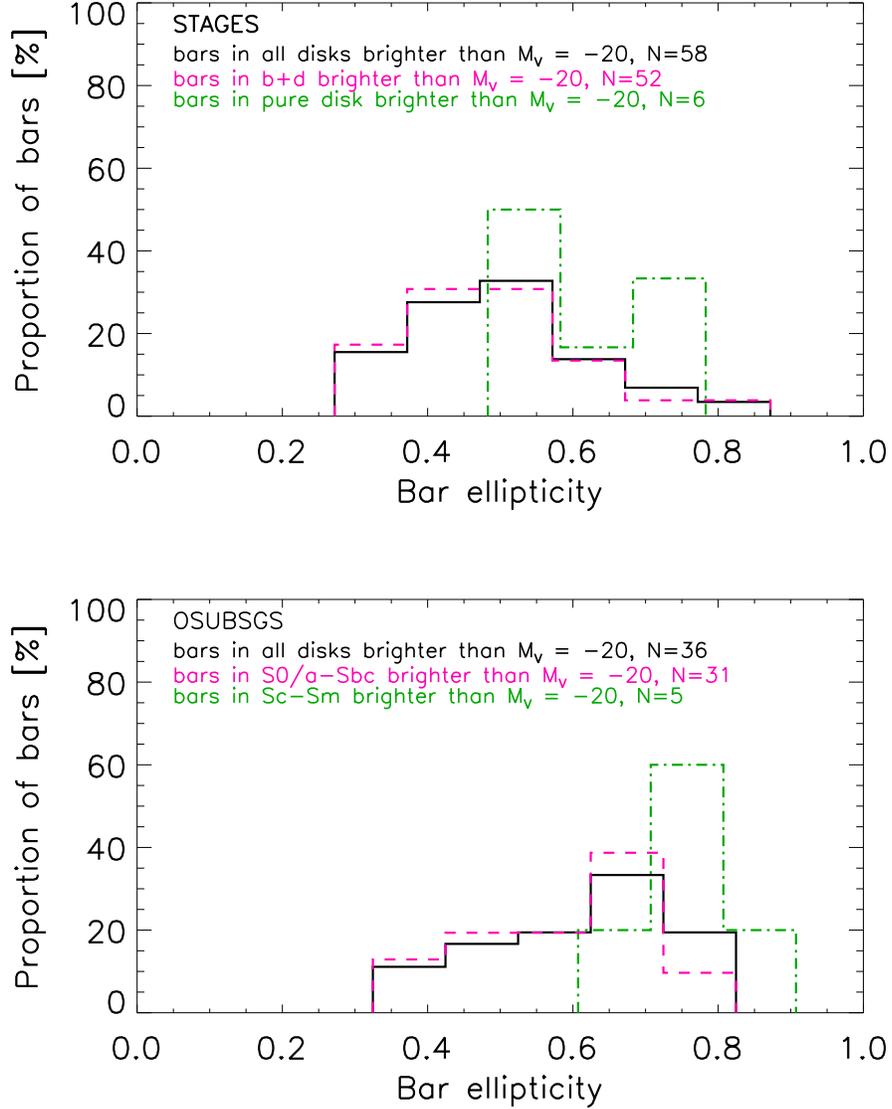}
\epsscale{0.7}
\vskip 1 in 
\caption{(a)~Distribution of bar peak ellipticity $e_{\rm bar}$
  for galaxies brighter than $M_{\rm V} = -20$ in the STAGES sample. The solid black line shows the ellipticity
  distribution for all bars. The pink and green lines show the
  ellipticity distributions for bars in galaxies visually classified
  as `bulge+disk' and `pure disk', respectively. In the STAGES sample, bars in galaxies
  classified as `bulge+disk' appear rounder than those in `pure disk'
  galaxies. (b)~Distribution of bar peak ellipticity $e_{\rm bar}$
  for galaxies brighter than $M_{\rm V} = -20$ in the OSUBSGS sample.  The pink and green lines show the
  ellipticity distributions for bulge-dominated (S0-Sbc) and
  disk-dominated (Sc-Sm) galaxies, respectively. 
  For both the STAGES and OSU samples, the strongest (highest
  ellipticity) bars are found in 
  disk-dominated, late-type galaxies. 
\label{ebosu}}
\end{figure}

\end{document}